\documentclass[aps,prd,twocolumn,amssymb,eqsecnum,showpacs,showkeyes,secnumarabic,graphics,floatfix,nofootinbib,tightenlines,longbibliography]{revtex4-1} 
 
\usepackage{graphicx}
\usepackage{epsfig}
\usepackage{bm}
\usepackage{cancel}
\usepackage{dcolumn}
\usepackage{amsmath}
\usepackage{hhline}
\usepackage{enumerate}
\usepackage[utf8]{inputenc}
\usepackage{cancel}
\usepackage[dvipsnames]{xcolor}
\usepackage{float}
\usepackage{longtable}
\usepackage[breaklinks=true,colorlinks=true,
linkcolor=blue,urlcolor=blue,citecolor=MidnightBlue,
bookmarks=true,bookmarksopenlevel=2]{hyperref}
\usepackage{mathrsfs}

\def \beq  {\begin{equation}}
\def \eeq  {\end{equation}}
\def \ber  {\begin{eqnarray}}
\def \eer  {\end{eqnarray}}

\begin{document}
\newcommand{\newc}{\newcommand}
\newc{\be}{\begin{equation}}
\newc{\ee}{\end{equation}}
\newc{\ba}{\begin{eqnarray}}
\newc{\ea}{\end{eqnarray}}
\newc{\bea}{\begin{eqnarray*}}
\newc{\eea}{\end{eqnarray*}}
\newc{\D}{\partial}
\newc{\ie}{{\it i.e.} }
\newc{\eg}{{\it e.g.} }
\newc{\etc}{{\it etc.} }
\newc{\etal}{{\it et al.}}
\newc{\lcdm}{$\Lambda$CDM }
\newc{\lcdmnospace}{$\Lambda$CDM}
\newcommand{\nn}{\nonumber}
\newc{\ra}{\Rightarrow}
\newc{\omm}{$\Omega_{m}$ }
\newc{\ommnospace}{$\Omega_{m}$}
\newc{\fs}{$f\sigma_8(z)$ }
\newc{\fsnospace}{$f\sigma_8(z)$}
\newc{\baodv}{$D_V(z) \times \frac{r_s^{fid}}{r_s}$}
\newc{\baodvspace}{$D_V(z) \times \frac{r_s^{fid}}{r_s}$ }
\newc{\baoh}{$H \times \frac{r_s}{r_s^{fid}}$}
\newc{\baohspace}{$H \times \frac{r_s}{r_s^{fid}}$ }
\newc{\baoda}{$D_A \times \frac{r_s^{fid}}{r_s}$}
\newc{\baodaspace}{$D_A \times \frac{r_s^{fid}}{r_s}$ }
\newc{\plcdm}{Planck15/$\Lambda$CDM }
\newc{\plcdmnospace}{Planck15/$\Lambda$CDM}
\newc{\wlcdm}{WMAP7/$\Lambda$CDM }
\newc{\wlcdmnospace}{WMAP7/$\Lambda$CDM}
\newcommand{\fss}{{\rm{\it f\sigma}}_8}

\title{Constraining power of cosmological observables: blind redshift spots and optimal ranges.}
\author{L. Kazantzidis}
\email{lkazantzi@cc.uoi.gr} 

\author{L. Perivolaropoulos}
\email{leandros@uoi.gr} 

\author{F. Skara}
\email{fskara@cc.uoi.gr} 

\affiliation{Department of Physics, University of Ioannina, 45110 Ioannina, Greece}

\date {\today}

\begin{abstract}
A cosmological observable measured in a range of redshifts can be used as a probe of a set of cosmological parameters. Given the cosmological observable and the cosmological parameter, there is an optimum range of redshifts where the observable can constrain the parameter in the most effective manner. For other redshift ranges the observable values may be degenerate with respect to the cosmological parameter values and thus inefficient in constraining the given parameter. These are blind redshift ranges. We determine the optimum and the blind redshift ranges of basic cosmological observables with respect to
three cosmological parameters: the matter density parameter \ommnospace, the equation of state parameter $w$ (assumed constant), and a modified gravity parameter $g_a$ which parametrizes a possible evolution of the effective Newton's constant as $G_{eff}(z)=G_N (1+g_a (1-a)^2 - g_a (1-a)^4)$ (where $a=\frac{1}{1+z}$ is the scale factor and $G_N$ is Newton's constant of General Relativity). We consider the following observables: the growth rate of matter density perturbations expressed through $f(z)$ and \fsnospace, the distance modulus $\mu(z)$, baryon acoustic oscillation observables \baodv, \baohspace and \baoda, $H(z)$ measurements and the gravitational wave luminosity distance. We introduce a new statistic $S_P^O(z)\equiv \frac{\Delta O}{\Delta P}(z) \cdot V_{eff}^{1/2}$, including the effective survey volume $V_{eff}$, as a measure of the constraining power of a given observable $O$ with respect to a cosmological parameter $P$ as a function of redshift $z$. We find blind redshift spots $z_b$ ($S_P^O(z_b)\simeq 0$) and optimal redshift spots $z_s$ ($S_P^O(z_s)\simeq max$) for the above observables with respect to the parameters \ommnospace, $w$ and $g_a$. For example for $O=f\sigma_8$ and $P=(\Omega_{m},w,g_a)$ we find blind spots at $z_b\simeq(1,2,2.7)$, respectively,  and optimal (sweet) spots at $z_s=(0.5,0.8,1.2)$. Thus probing higher redshifts may in some cases be less effective than probing lower redshifts with higher accuracy. These results may be helpful in the proper design of upcoming missions aimed at measuring cosmological obsrevables in specific redshift ranges.  
\end{abstract}  
\maketitle

\section{Introduction}
\label{sec:Introduction}

The validity of the standard cosmological model (\lcdm \cite{Carroll:2000fy}) is currently under intense investigation using a wide range of cosmological observational probes including Cosmic Microwave Background (CMB) experiments,  galaxy photometric and spectroscopic surveys, attempts to measure Baryon Acoustic Oscillations (BAO), Weak Lensing (WL), Reshift Space Distortions (RSD), cluster counts, as well as the use of Type Ia Supernovae (SnIa) as standard candles. 

This investigation has revealed the presence of tensions within the \lcdm model, i.e. inconsistencies among the parameter values determined using different observational probes. The most prominent tension is the $H_0$ tension which indicates $3\sigma$ level inconsistencies between the value favored by the latest CMB data release  from the Planck Collaboration \cite{planck15,planck18}  
[$H_0=67.4\pm 0.5 km s^{-1} Mpc^{-1}$ (68\% confidence limit)] and the local Hubble Space Telescope measurement \cite{Riess:2016jrr} (based on distance ladder estimates from Cepheids) $H_0=73.24\pm 1.74 km s^{-1} Mpc^{-1}$ (68\% confidence limit). Another less prominent tension ($2-3\sigma$) is the $\Omega_{m}-\sigma_8$ tension between the CMB Planck data and the growth of density perturbations data (RSD and WL) \cite{Macaulay:2013swa,Kazantzidis:2018rnb,Nesseris:2017vor,kids1,kids2}. The CMB data favor higher values of the matter density parameter $\Omega_{m}$ and the matter fluctuations amplitude $\sigma_8$ than the data that probe directly the gravitational interaction (RSD and WL). 

A key question therefore arises: {\it Are these tensions an early hint of new physics beyond the standard model or are they a result of systematic/statistical fluctuations in the data?}

Completed, ongoing and future CMB experiments and large scale structure surveys aim at testing the standard \lcdm model and addressing the above question. These surveys are classified in four stages. Stages I and II correspond to completed surveys and CMB experiments, while stages III and IV correspond to ongoing and future projects respectively. For example stage II CMB experiments include WMAP \cite{wmap1}, 
Planck \cite{planck15,planck18}, ACTPol \cite{actpol} and  SPT-Pol \cite{sptpol}, while stage III CMB experiments include AdvACT \cite{advact} and SPT-3G \cite{spt3g}. Future stage IV CMB probes on the ground\cite{cmbgroundiv} and in space such as LiteBIRD \cite{litebird,litebird1} mainly aim to measure CMB lensing and the CMB-B modes in detail.

A large amount of high-quality data is expected in the coming years from large scale structure surveys (see Table \ref{table:surveys}). Stage III large scale structure surveys include the Canada-France-Hawaii Telescope Lensing Survey \cite{cfhtlens}, the Kilo Degree Survey (KiDS) \cite{kids1,kids2}, the extended Baryon Oscillation Spectroscopic Survey (eBOSS) \cite{eboss}, the Dark Energy Survey (DES) \cite{des1,des2,des3} and the Hobby Eberly Telescope Dark Energy Experiment (HETDEX) \cite{hetdex}. Finally, stage IV large scale structure surveys include ground-based telescopes such as the Dark Energy Spectroscopic Instrument (DESI), the Large Synoptic Survey Telescope (LSST) \cite{lsst1,lsst2} and the Square Kilometer Array (SKA) \cite{ska1,ska2,ska3,ska4} as well as space based telescopes such as Euclid \cite{euclid1,euclid2} and the Wide Field Infrared Survey Telescope (WFIRST) \cite{wfirst1,wfirst2}. The redshift ranges of these and other similar surveys are shown along with their type and duration in Table \ref{table:surveys}.
\begin{widetext} 
\begin{center}
\begin{table}[H]
\centering
\begin{tabular}{|c|c|c|c|c|}
\hline 
{\bf Survey} & $z${\bf  Range} & {\bf Type} &{\bf Duration} & {\bf Ref.} \tabularnewline
\hline 
\hline 
SDSS & $0.1<z<0.6$ &Spectroscopic & 2006-2010 &\cite{sdss} \tabularnewline
\hline 
WIGGLEZ & $0.4<z<0.8$ &Spectroscopic & 2006-2010 &\cite{boss} \tabularnewline
\hline 
BOSS & $0.35$, $0.6$, $2.5$ &Spectroscopic & 2009-2014 &\cite{boss} \tabularnewline
\hline 
KIDS & $0<z<0.8$ &Photometric & 2011-  &\cite{kids1,kids2} \tabularnewline
\hline 
DES & $0.3<z<1.0$ &Photometric & 2012-2018 &\cite{des1,des2,des3} \tabularnewline
\hline
HETDEX & $1.9<z<3.5$ &Spectroscopic & 2015-2017 &\cite{hetdex} \tabularnewline
\hline
eBOSS & $0.6<z<2.2$ &Spectroscopic & 2015-2018 &\cite{eboss} \tabularnewline
\hline   
DESI & $0.6<z<1.7$ &Spectroscopic & $>2019$ &\cite{desi1,desi2,desi3} \tabularnewline
\hline 
DESI-Bright Galaxies & $0.0<z<0.4$ &Spectroscopic & $>2019$ &\cite{desi1,desi2,desi3} \tabularnewline
\hline 
Euclid & $0.8<z<2.0$ &Spectroscopic & 2022-2027 &\cite{euclid1,euclid2,euclid3}
\tabularnewline
\hline 
LSST & $0.5<z<3$ &Photometric & $>2019$ &\cite{lsst1,lsst2}
\tabularnewline
\hline
WFIRST & $1<z<3$ &Spectroscopic & $>2020$ &\cite{wfirst1,wfirst2}
\tabularnewline
\hline
\end{tabular}
\caption{Some recent and future large-scale structure surveys. Photometric surveys focus mainly on WL, while spectroscopic surveys measure mainly RSD. The redshift range shifts to higher redshifts for stage III and stage IV surveys.}
\label{table:surveys}
\end{table}
\end{center}
\end{widetext}
As seen in Table \ref{table:surveys}, the redshift ranges of more recent surveys tend to increase in comparison with earlier surveys. This trend for higher redshifts implies an assumption of increasing constraining power of observables on cosmological parameters with redshift. As demonstrated in the present analysis however, this assumption is not always true. In this context the following questions arise:
\begin{enumerate}[{(1)}]
\item
What is the redshift dependence of the constraining power of a given observable with respect to a given cosmological parameter?
\item
Is there an optimal redshift range where the constraining power of a given observable is maximal with respect to a given cosmological parameter?
\item 
Are there blind redshift spots where a given observable is degenerate with respect to specific cosmological parameters?
\end{enumerate}
These questions are addressed in the present analysis. Previous studies \cite{Nesseris:2011pc}
have indicated the presence of degeneracies for the case of growth of fluctuations observable \fs with respect to the equation of state parameter $w$ in specific redshift ranges. Here, we extend these results to a wider range of observables and cosmological parameters. 

In particular the goals of the present analysis are the following
\begin{enumerate}[{(1)}]
\item
Present extensive up-to-date compilations of recent measurements of cosmological observables including growth of perturbations, BAO, and luminisity distance  observables. 
\item
Identify the sensitivity of these observables as a function of redshift for three cosmological parameters: the present matter density parameter \ommnospace, the dark energy equation of state parameter $w$ (assumed constant), and a parameter $g_a$ describing the evolution of the effective Newton's constant in the context of a well motivated parametrization \cite{Nesseris:2017vor,Kazantzidis:2018rnb}.
\item
Identify possible trends for deviations of the above parameters from their standard \plcdm values in the context of the above data compilations.
\end{enumerate}

The structure of this paper is as follows. In the next section we review  the basic equations determining the growth of cosmological density perturbations. These equations can lead to the predicted evolution of the observable product $f(a)\sigma_8(a)$, where $a$ is the scale factor $a=\frac{1}{1+z}$,  $f(a)\equiv d\ln \delta(a)/d\ln a$ is the growth rate of cosmological perturbations, $\delta(a)\equiv \delta \rho/\rho$ is the linear matter overdensity growth factor, and $\sigma_8$ is the matter power spectrum normalisation on scales of $8 h^{-1} Mpc$. In this section we discuss the sensitivity of the observables $f\sigma_8(z)$ and $f(z)$ on the cosmological parameters \ommnospace, $w$ and $g_a$ as a function of redshift. The redshift range of the current available data \fs that is most constraining on these parameters is also identified and the existence of blind redshift spots where \fs is insensitive to these parameters is demonstrated. The selection of these particular parameters (\ommnospace, $w$ and $g_a$) is important as their combination can lead to direct test of General Relativity (GR) by simultaneously constraining the background expansion rate through $H(z)$ and the  possible evolution of the effective Newton’s constant.  It is important to notice that the evolution of the effective Newton’s  $G_{eff}(z)$ obtained through the parameter $g_a$ is degenerate with $H(z)$ constant and can only be probed once $H(z)$  is also efficiently constrained through the parameters $\Omega_{m}$ and  $w$. 

In Sec. \ref{sec:BAO} we focus on cosmological observables obtained from BAO data, present an updated extensive compilation of such data, and identify the sensitivity of the BAO observables on the parameters \ommnospace, $w$ and $g_a$ as a function of redshift. As in the case of the growth observables, blind redshift spots and optimal redshift ranges are identified. The effects of the data redshift range on the shape and size of the uncertainty contours in the above cosmological parameter space are also identified. In Sec. \ref{sec:Distance Moduli} we focus on luminosity distance moduli as obtained from type Ia supernovae and gravitational waves and identify the sensitivity 
of these observables to the parameters \ommnospace, $w$ and $g_a$ as a function of redshift. Binned JLA data are superimposed on the plots to demonstrate the sensitivity of the distance moduli to the cosmological parameters. Finally in Sec. \ref{sec:Discussion} we conclude, summarize and discuss future prospects of the present analysis.  

\section{Growth of Density Perturbations: The Observables \fs and $f(z)$}
\begin{figure*}
\begin{centering}
\includegraphics[width =0.98\textwidth]{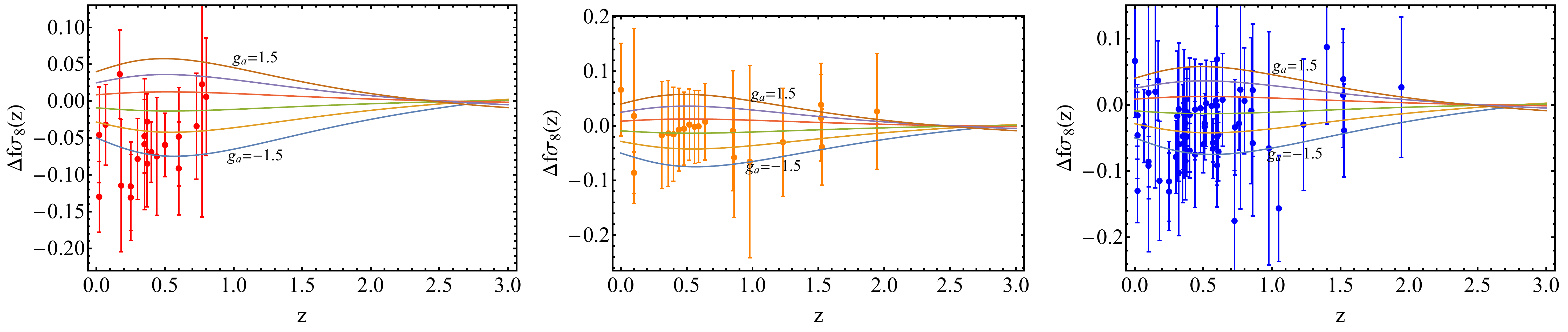}
\par\end{centering}
\caption{$\Delta f \sigma_8$ as a function of redshift for various values of $g_\alpha$ superimposed with the early growth data (left panel), late data (middle panel) and full growth data (right panel).}  
\label{fig:delfs8ga}
\end{figure*}

The evolution of the linear matter density growth factor $\delta\equiv \delta \rho/\rho$ in the context of both GR and most modified gravity theories on subhorizon scales is described by the equation 
\be
{\ddot \delta} + 2H {\dot \delta} - 4\pi G_{\rm eff}\,
\rho \, \delta \approx 0 \label{eq:odedeltat}
\ee
where $\rho$ is the background matter density and  $G_{\rm eff}$ is the effective Newton's constant (which in general depends on redshift $z$ and cosmological scale $k$), and $H$ is the Hubble parameter. In terms of the redshift $z$,  Eq. \eqref{eq:odedeltat} takes the form 
\begin{widetext}
\begin{equation}
\delta'' + \left(\frac{(H(z)^2)'}{2~H(z)^2} -
{1\over 1+z}\right)\delta'
-{3\over 2} \frac{(1+z)~\Omega_{m}~ G_{eff}(z,k)/
G_{N}}{H(z)^2/H_0^2}~\delta=0
\label{eq:odedeltaz}
\end{equation}
\end{widetext}
while in terms of the scale factor we have
\be
\delta''(a)+\left(\frac{3}{a}+\frac{H'(a)}{H(a)}\right)\delta'(a)
-\frac{3}{2}
\frac{\Omega_{m} G_{eff}(a,k)/G_{N}}
{a^5 H(a)^2/H_0^2} 
\delta(a)=0 \label{eq:odedeltaa}
\ee
$G_{eff}$ arises from a generalized Poisson equation 
\be 
\nabla^2 \phi \approx 4 \pi G_{\rm eff} \rho \; \delta
\ee
where $\phi$ is the perturbed metric potential in the Newtonian gauge where the perturbed FRW metric takes the form
\be
ds^2= -(1 + 2 \phi) dt^2 + a^2 (1 - 2\psi) d{\vec{x}}\,^2
\ee
GR predicts a constant homogeneous $G_{\rm eff}(z,k)=G_N$ ($G_N$ is Newton's constant as measured by local experiments) 

Constraints from Solar System \cite{Nesseris:2006hp}  and nucleosynthesis tests \cite{Copi:2003xd} imply that $G_{\rm eff}$ is close to the GR predicted form in both low and high redshifts. In particular at low $z$ we have \cite{Nesseris:2006hp} 
\be 
\Big\lvert \frac{1}{G_N} \frac{d G_{\rm eff}(z)}{dz}\vert_{z=0} \Big\rvert < 10^{-3} h^{-1}
\label{eq:geffconstr1}
\ee
while the second derivative is effectively unconstrained since 
\be 
\Big\vert \frac{1}{G_N}  \frac{d^2G_{\rm eff}(z)}{dz^2}\vert_{z=0}\Big\vert < 10^{5} h^{-2}
\label{eq:geffconstr2}
\ee
At high $z$ \cite{Copi:2003xd} and at $1\sigma$, we have
\be
\lvert G_{\rm eff}/G_{N} -1 \rvert \leq 0.2
\label{eq:geffnucconstr}
\ee

A parametrization of $G_{\rm eff}(z)$ respecting these constraints is of the form \cite{Nesseris:2017vor}
\ba
\frac{G_{\textrm{eff}}(a,g_a,n)}{G_{\textrm{N}}} &=& 1+g_a(1-a)^n - g_a(1-a)^{n+m} \nn \\
&=&1+g_a\left(\frac{z}{1+z}\right)^n - g_a\left(\frac{z}{1+z}\right)^{n+m}. \label{eq:geffansatz}
\ea
where $n$ and $m$ are integer parameters with $n\geq 2$ and $m>0$. Here we set $n=m=2$. 

The observable $f\sigma_8(a)$ can be obtained from the solution $\delta(a)$ of Eq. \eqref{eq:odedeltaa} using the definitions $f(a)\equiv d\ln \delta(a)/ d\ln a$ and $\sigma(a)=\sigma_8 \frac{\delta(a)}{\delta(1)}$. Thus, we have \cite{Percival:2005vm}
\ba
\fss(a)&\equiv& f(a)\cdot \sigma(a)\\
&=&\frac{\sigma_8}{\delta(1)}~a~\delta'(a) ,\label{eq:fs8}
\ea

Therefore, both $f\sigma_8(a)$ and the growth rate $f(a)$ [or equivalently $f\sigma_8(z)$ and $f(z$)] can be obtained by numerically solving Eq. \eqref{eq:odedeltaz} or \eqref{eq:odedeltaa}. The solution of these equations requires the specification of proper parametrizations for both the background expansion $H(z)$ and the effective Newton's constant $G_{eff}(z)$. In the context of the present analysis we assume a flat universe and a $wCDM$ model background expansion of the form 
\begin{gather} 
H^2(z)=H_0^2 \left[\Omega_{m}(1+z)^3 +(1-\Omega_{m})(1+z)^{3(1+w)}\right] \Rightarrow \nonumber \\
E^2(z)=\frac{H^2(z)}{H_0^2}=\Omega_{m}(1+z)^3 +(1-\Omega_{m})(1+z)^{3(1+w)}
\label{eq:hzwcdm}
\end{gather}
and $G_{eff}$ parametrized by Eq. \eqref{eq:geffansatz} with $n=m=2$. Using these parametrizations and initial conditions corresponding to GR in the matter era [$\delta(a)\sim a$] it is straightforward to obtain the predicted evolution of the observables \fs and $f(z)$ for various parameter values around the standard \plcdm model parameters ($\Omega_{m}^P=0.31$, $w=-1$, $g_a=0$). For each observable $O(\Omega_{m},w,g_a)$ [e.g. \fsnospace] we consider the deviation\footnote{In certain cases we consider the deviation around \omm=0.3 instead of \omm=$\Omega_m^P$}  
\be
\Delta O_{\Omega_{m}} \equiv O(\Omega_{m},-1,0)-O(\Omega_{m}^P,-1,0) \label{eq:devdef}   
\ee
Similar deviations $\Delta O_w$ and $\Delta O_{g_a}$ are defined  for the other two parameters in the context of a given observable $O$.

In Fig. \ref{fig:delfs8ga} we show the deviation $\Delta f\sigma_{8g_a}$ for $g_a$ in the range $g_a\in[-1.5,1.5]$ superposed with a recent compilation of the \fs data \cite{Kazantzidis:2018rnb} shown in Table \ref{tab:data-rsd} in the Appendix (with early data published before 2015 in the left panel, recent data published after 2016 in the middle panel and full dataset in the right panel). No fiducial model correction has been implemented for the datapoints shown, but such a correction would lead to a change of no more than about  $3\%$ \cite{Macaulay:2013swa,Kazantzidis:2018rnb}.
There are three interesting points to be noted in Fig. \ref{fig:delfs8ga}.
\begin{enumerate}[{(1)}]
\item
Early data favor weaker gravity ($g_a<0$) for redshifts around $z\simeq 0.5$ assuming a fixed \plcdm background. This trend is well known \cite{Macaulay:2013swa} and has been demonstrated and discussed extensively, e.g. in Refs. \cite{Nesseris:2017vor,Gomez-Valent:2017idt,Gomez-Valent:2018nib,Basilakos:2017rgc,Basilakos:2016nyg,Alam:2015rsa,LHuillier:2017ani,Sagredo:2018ahx,Arjona:2018jhh}.
\item
The observable \fs has a blind spot with respect to the parameter $g_a$ at redshift $z\simeq 2.7$. Such a blind spot was also pointed out in Ref. \cite{Nesseris:2011pc} with respect to a similar gravitational strength parameter (where it was called ``sweet spot" in that Ref. \cite{Nesseris:2011pc} even though the term ``blind spot" should have been used).
\item 
There is a redshift range around $z\simeq 0.5$ of optimal sensitivity of the observable \fs with respect to the parameter $g_a$. Despite of the existence of this optimal redshift range much of the recent \fs data appear at larger redshifts approaching the blind spot region. These datapoints have reduced sensitivity in identifying deviations of $G_{eff}$ from its GR value $G_N$
\end{enumerate}

The existence of blind spots and optimal redshifts of an observable $O$ with respect to a cosmological parameter $P$ may also be quantified by defining the ``sensitivity" measure including the effects of the survey volume $V_{eff} (k,z)$. The effective survey volume probed for a particular k mode with the power spectrum $p(k,z)$ in a survey of sky area surveyed $\Delta\Omega$ is given by
\be
V_{eff}(k,z)=\Delta\Omega\int_0^{z} \left[\frac{n(z')p(k,z')}{1+n(z')p(k,z')} \right]^2\frac{dV}{dz' d\Omega} dz' 
\ee
where $z$ is the maximum redshift corresponding to the survey volume $V_{eff}$ and $n(z)$ is the number density of galaxies that are detected, which is given as    
 \be
n(z)=\int_{M_{lim}(z)}^{\infty}\frac{dN}{dV dM}dM  
 \ee
The function $M_{lim}(z)$ is the limiting mass threshold which is detected for the given survey 
and $dV$ is the infinitesimal comoving volume element
\be
 dV=\frac{r^2(z)}{H(z)} d\Omega \, dz  
\ee 
where 
\be 
r(z)= \frac{c}{H_0}\int_0^z\frac{dz'}{E(z')}   
\ee
and $E(z')$ is given by Eq. \eqref{eq:hzwcdm}

The constraining power of the observable $O$ depends on the survey volume  $V_{eff} (k,z)$, since the error $\sigma_p$ on the measurement of the power spectrum $p(k,z)$ increases as the effective survey volume  $V_{eff} (k,z)$ decreases (i.e. as less $k$ modes are measured by the survey) as \cite{Feldman:1993ky,Huelga:1997mw,Duffy:2014lva,Abdalla:2004ah} \\
\be
\left(\frac{\sigma_p}{p(k,z)}\right)^2=\frac{2}{4\pi k^3\Delta(log \, k)}\frac{(2\pi)^3}{V_{eff}(k,z)} \left[\frac{1+n(z)p(k,z)}{n(z) p(k,z)} \right]^2 \label{errorsigmapi}   
\ee

Thus, since the error $\sigma_p$ on the measurement of the power spectrum $p(k,z)$ is inversely  proportional to the square root of he survey volume $V_{eff}(k,z)$ [see  Eq. \eqref{errorsigmapi}], we define the ``sensitivity" measure  as
\be  
S_P^O\equiv \frac{\Delta O(P)}{\Delta P} \cdot V_{eff}(k,z)^{1/2}  
\label{sensdef}
\ee
where $\Delta O$ is the deviation of the observable $O$ when a given parameter varies in a fixed small range  $\Delta P=P_{max}-P_{min}$ around a fiducial model value (e.g. \plcdmnospace). 
In Fig.  \ref{fig:sensdiffparmeff} we show a plot of the sensitivity measure $S$ for the observable \fs and the three parameters \ommnospace, $w$, $g_a$. The existence of blind spots is manifest as roots of the sensitivity measure, while optimal redshifts appear as maxima of the magnitude of $S$. We have fixed $k$ such as that $np=3$ assuming sufficient signal to noise per pixel \cite{Duffy:2014lva}. We have also rescaled  sensitivity measure statistic so that it is unity at its maximum absolute value. The nonlinear modes may be excluded by setting a minimum redshift which is of $O(10^{-2} )$ and are much smaller than the derived optimal redshifts and blind spots identified in our analysis. Notice that the sensitivity measure indicates the presence of blind spots for all three parameters. For $w$ the blind spot is close to $z\simeq 2$ while for \omm is close to $z\simeq 1$. The corresponding optimal redshifts are  at  $z\simeq 1.2$ for $g_a$, at $z\simeq 0.8$ for $w$ and at $z\simeq 0.5$ for $\Omega_m$. (Although the region $z>2$ for $w$ and $\Omega_m$ provides better sensitivity, there are currently  almost no data available in this redshift range). Notice also in Figs. \ref{fig:delfs8ga} and \ref{fig:sensdiffparmeff} that when including the effects of the survey volume the optimal redshifts shift to somewhat higher redshifts, while the blind spots remain unaffected.

As shown in Figs. \ref{fig:delfs8w} and \ref{fig:delfs8om} for both cases, recent data approach the blind spot regions in contrast to early published data that efficiently probed the optimal redshift regions for both parameters $w$ and \ommnospace. Also, early data seem to favor weaker growth of perturbations which occurs for lower, $g_a$, and \omm and higher $w$ \cite{Macaulay:2013swa,Nesseris:2017vor,Kazantzidis:2018rnb}. If this trend is partly attributed to a lower value of $G_{eff}$ in the recent past, then it is difficult to reconcile with the most generic modified gravity theories like $f(R)$ and scalar tensor theories \cite{Nesseris:2017vor,Gannouji:2018ncm} 
\begin{widetext}
\begin{figure*}
\begin{centering}
\includegraphics[width =0.98\textwidth]{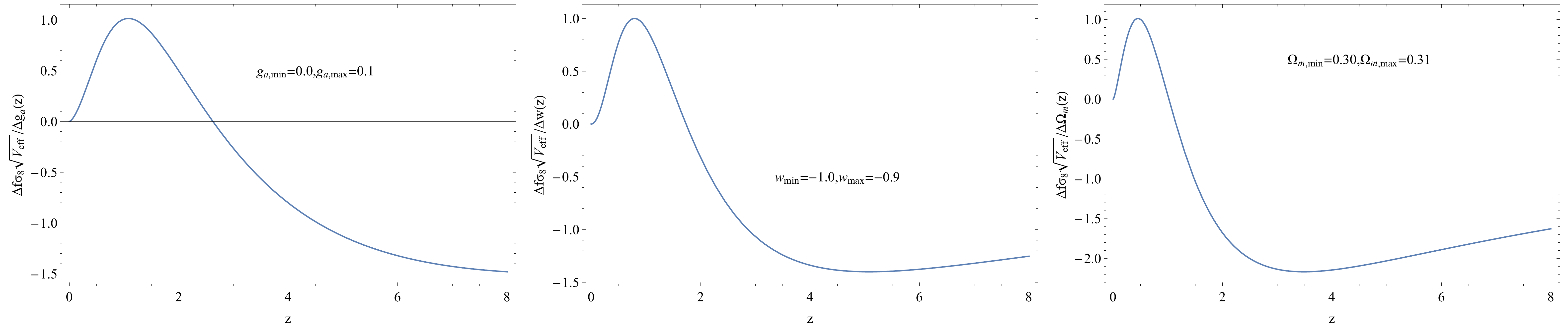}
\par\end{centering}
\caption{The sensitivity measure of $\frac{\Delta f \sigma_8}{\Delta P}V_{eff}^{1/2}$ for $P=g_\alpha$ (left panel), $P=w$ (middle panel), and $P=\Omega_m$ (right panel) }  
\label{fig:sensdiffparmeff} 
\end{figure*}

\begin{figure*} 
\begin{centering}
\includegraphics[width =0.98\textwidth]{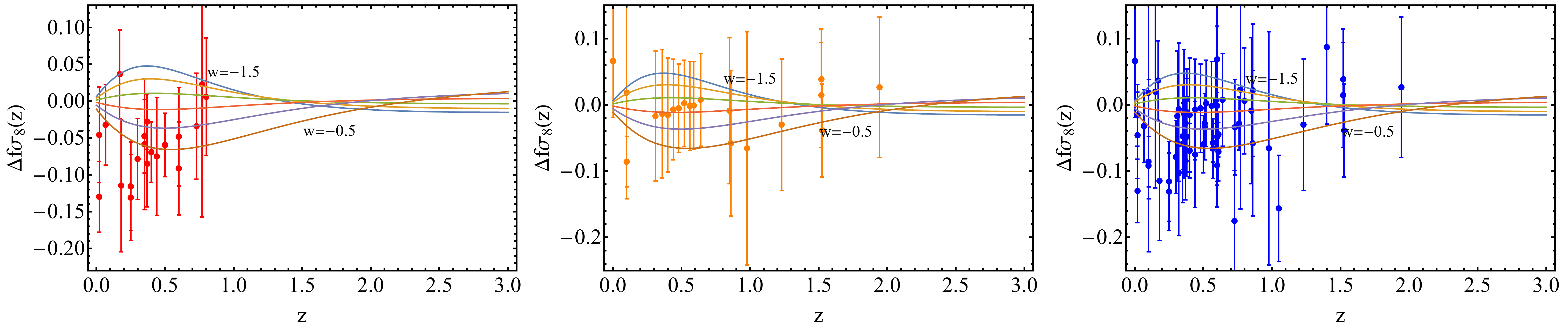}
\par\end{centering}
\caption{Same as Fig. \ref{fig:delfs8ga} for various values of $w$.}  
\label{fig:delfs8w}
\end{figure*}

\begin{figure*}
\begin{centering}
\includegraphics[width =0.98\textwidth]{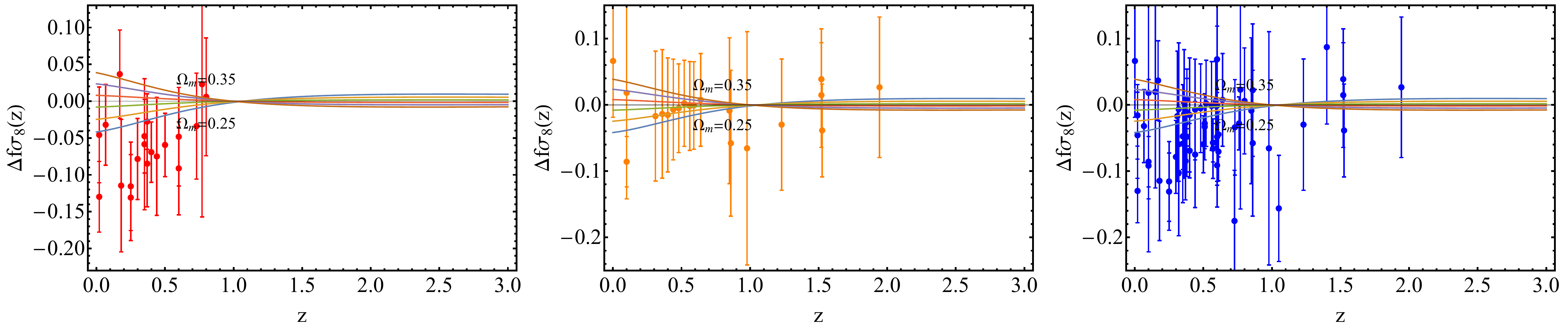}
\par\end{centering}
\caption{Same as Fig. \ref{fig:delfs8ga} for various values of \ommnospace.}  
\label{fig:delfs8om}
\end{figure*}
\end{widetext}

\begin{figure*}
\begin{centering}
\includegraphics[width =0.95\textwidth]{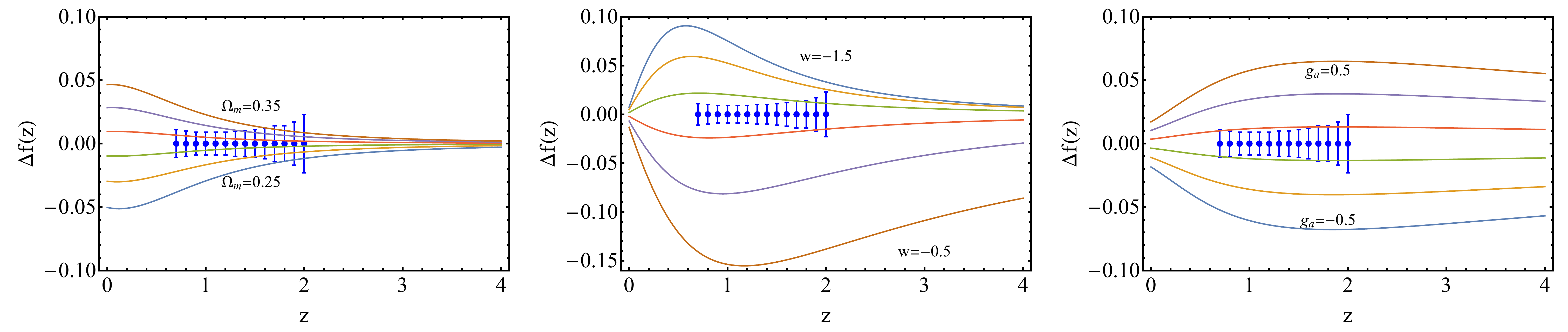}
\par\end{centering}
\caption{$\Delta f(z)$ as a function of redshift superimposed with the Euclid mock data for different values of \omm (left panel), $w$ (middle panel), and $g_\alpha$ (right panel).}  
\label{fig:deltfsall}
\end{figure*}

A similar analysis can be performed for the growth rate observable $f(z)$ which will be probed by the Euclid mission \cite{euclid2}. Mock Euclid data assuming a \plcdm fiducial model are shown in Fig. \ref{fig:deltfsall} with proper redshifts and error bars \cite{euclid2} along with the deviation of  the observable $f(z)$ with respect to \omm (left panel), $w$ (middle panel) and $g_a$ (right panel). Clearly, the predicted redshift range of the Euclid data is optimal for the identification of new gravitational physics (right panel), but it is not optimized for constraining the matter density parameter (left panel of Fig.  \ref{fig:deltfsall}) or the equation of state parameter if $w<-1$ (middle panel). 

The observable $f(z)$ is considered due to the approach of Ref. \cite{euclid2}, where the Euclid team indicated that the large number of galaxies  of the Euclid survey combined with the depth of the survey will allow a reliable estimate of the bias  simultaneously with the growth rate $f(z)$ obtained through the redshift distortion $\beta$. The redshift distortion $\beta$ is defined as
\be
\beta(z)=\frac{\Omega_m(z)^{\gamma}}{b(z)}= \frac{f(z)}{b(z)}
\ee
where $b(z)$ is the bias. Thus, the survey will not only probe the bias-free combination $f\sigma_8$, but also directly probe the growth observable $f(z)$ which is modeled in Ref. \cite{euclid2} with errorbars and is also considered separately in our analysis. Of course, what is actually observable is the redshift distortion $\beta$ parameter which is obtained through the ratio between the monopoles of the correlation functions in real and in redshift space. Thus, the derived blind spot and optimal redshift for the growth rate $f(z)$ are accurate under the assumption that the bias $b(z)$ has a very weak dependence on the redshift.

\section{Baryon Acoustic Oscillations: the Observables \baodv, \baohspace and \baoda }
\label{sec:BAO}
\subsection{BAO Observables and their Variation with Cosmological Parameters.}
Waves induced by radiation pressure in the pre-recombination plasma inflict a characteristic BAO scale on the late-time matter clustering at the radius of the sound horizon, defined as
\be 
r_s =\int_{z_{d}}^\infty \frac{c_s(z)}{H(z)} dz 
\label{rsdef}
\ee
where  $c_s$ is given by \cite{Komatsu:2008hk} 

\be 
c_s(z)=\frac{c}{\sqrt{3 \left(1+\frac{3\Omega_b}{4\Omega_\gamma} \frac{1}{1+z}\right)}} \label{csdef}   
\ee
and the drag redshift $z_d$ corresponds to times shortly after recombination, when photons decouple from baryons \cite{Aubourg:2014yra}. This BAO scale appears as a  peak in the correlation function or equivalently as damped  oscillations in the large scale structure power spectrum. In the context of standard matter and radiation epochs, the Planck 2015 measurements of the matter and baryon densities $\Omega_m$ and $\Omega_b$ specify the BAO scale to great accuracy (uncertainty less than $1\%$). An anisotropic BAO analysis
measuring the sound horizon scale along the line of sight and along the transverse direction can  measure both $H(z)$ and the comoving angular
diameter distance $D_M(z)$ related to the physical angular
diameter distance in  a flat universe

\be
D_A(z)=\frac{1}{1+z}\int_0^z\frac{c\; dz}{H(z)}
\label{dazdef}
\ee
as $D_M(z) = (1 + z)D_A(z)$  \cite{Bautista:2017zgn}. Deviation of cosmological parameters can change $r_s$, so BAO measurements actually constrain
the combinations $D_M(z)\times\frac{r_s^{fid}}{r_s}$ or equivalently $D_A(z)\times\frac{r_s^{fid}}{r_s}$, $H(z)\times\frac {r_s}{r_s^{fid}}$ where $r_s^{fid}$ is the sound horizon (BAO scale) in the context of the fiducial cosmology assumed in the construction of the large-scale structure correlation function. An angle-averaged galaxy
BAO measurement constrains the  combination 
\be  
D_V(z)=\left[czD_M(z)^2/H(z)\right]^{1/3} 
\label{dvdef}
\ee
Taking into account the variation of cosmological parameters the constrained combination becomes \baodv.
Statistical isotropy can be used to constrain the observable combination $H(z)D_M(z)$ using an anisotropic BAO analysis in the context of the Alcock-Paczynski test \cite{Alcock:1979mp}.
The sound horizon $r_s(z_d)$ at the drag epoch $z_d$ that enters the BAO observables  may be calculated in the context of a given cosmological model, either numerically (e.g. with CAMB \cite{Lewis:1999bs}) or using a fitting formula for $z_d$ \cite{Eisenstein:1997ik} of the form 
\begin{equation}
z_d  =
 \frac{1291(\Omega_{m}h^2)^{0.251}}{1+0.659(\Omega_{m}h^2)^{0.828}}
\left[1+b_1(\Omega_bh^2)^{b_2}\right]
\label{eq:zd}
\end{equation}
where
\begin{eqnarray}
  b_1&=&0.313(\Omega_{m}h^2)^{-0.419}\left[1+0.607(\Omega_{m}h^2)^{0.674}\right]\\
  b_2&=&0.238(\Omega_{m}h^2)^{0.223}
\end{eqnarray}
and from Eq. \eqref{rsdef}
\begin{equation}
 r_s(z)
=\frac{c}{\sqrt{3}}\int_{z_d}^{\infty}
\frac{dz}{H(z)\sqrt{1+\frac{3\Omega_b}{4\Omega_\gamma} \frac{1}{1+z}}}  
\label{eq:rs}
\end{equation} 

where $\Omega_\gamma=2.469\times 10^{-5}h^{-2}$ for $T_{\rm
cmb}=2.725$~K, and
\begin{equation}
 H(z) = H_0\left[\Omega_m (1+z)^3+\Omega_r (1+z)^4+ \Omega_\Lambda (1+z)^{3(1+w)}\right]^{1/2}
\label{eq:hubble}
\end{equation}
with $\Omega_r=\Omega_\gamma (1+0.2271 N_{eff})$ ($N_{eff}\simeq 3$ is the number of neutrino species) and 

\begin{equation}
 \Omega_m + \Omega_r + \Omega_\Lambda = 1
\end{equation}
in the context of a flat universe. It has been shown \cite{Hamann:2010pw} that when the fitting formula is used to obtain $z_d$ close to the \plcdm parameter values, a correction factor of $154.66/150.82$ should be used on $r_s$ obtained from Eq. \eqref{eq:rs} to obtain agreement with the more accurate numerical estimate of $r_s$. Using Eqs. \eqref{dazdef}, \eqref{dvdef}, \eqref{eq:rs} and a \plcdm fiducial cosmology ($h=0.676$, $\Omega_b h^2=0.0223$, $\Omega_{m}=0.31$ and $r_s^{fid}=147.49$ Mpc), it is straightforward to construct the theoretically predicted redshift dependence of the BAO observables \baodv, \baohspace and \baodaspace for various values of the parameters \omm and $w$ and superpose this dependence with corresponding currently available data shown in Table \ref{tab:data-bao} in the Appendix.
\begin{widetext}
\begin{figure*}
\begin{centering}
\includegraphics[width =0.95\textwidth]{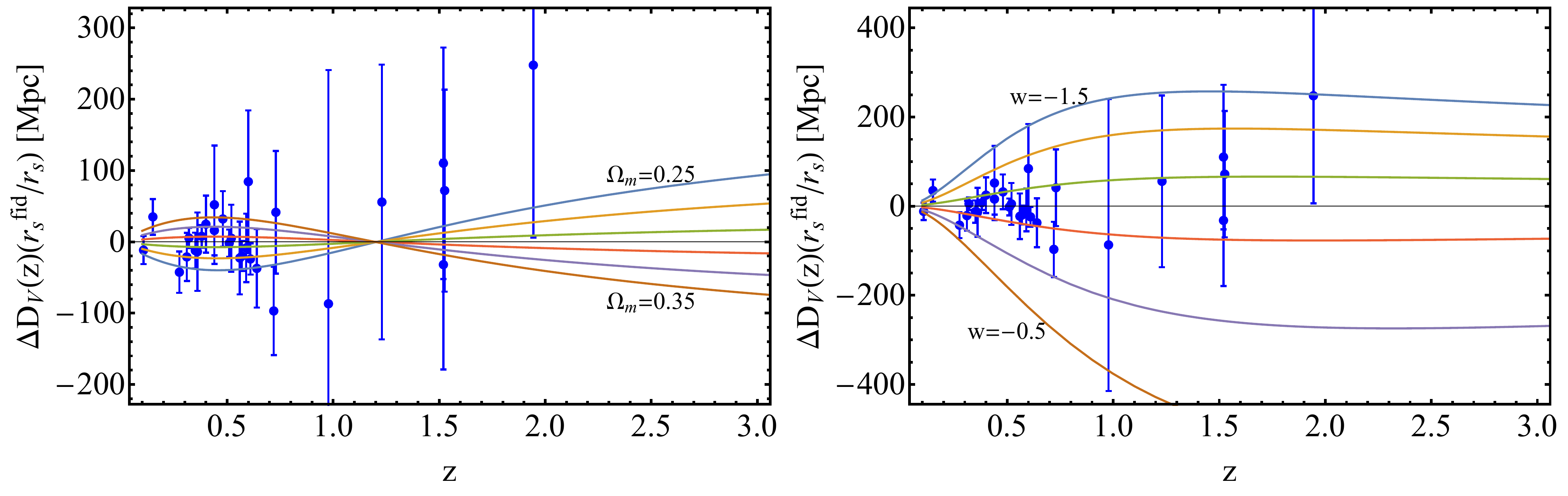}
\par\end{centering}
\caption{The deviation $\Delta$\baodvspace as a function of the redshift $z$ for different values of \omm (left panel) and $w$ (right panel).}  
\label{fig:dvfig}
\end{figure*}
\begin{figure*}
\begin{centering}
\includegraphics[width =0.95\textwidth]{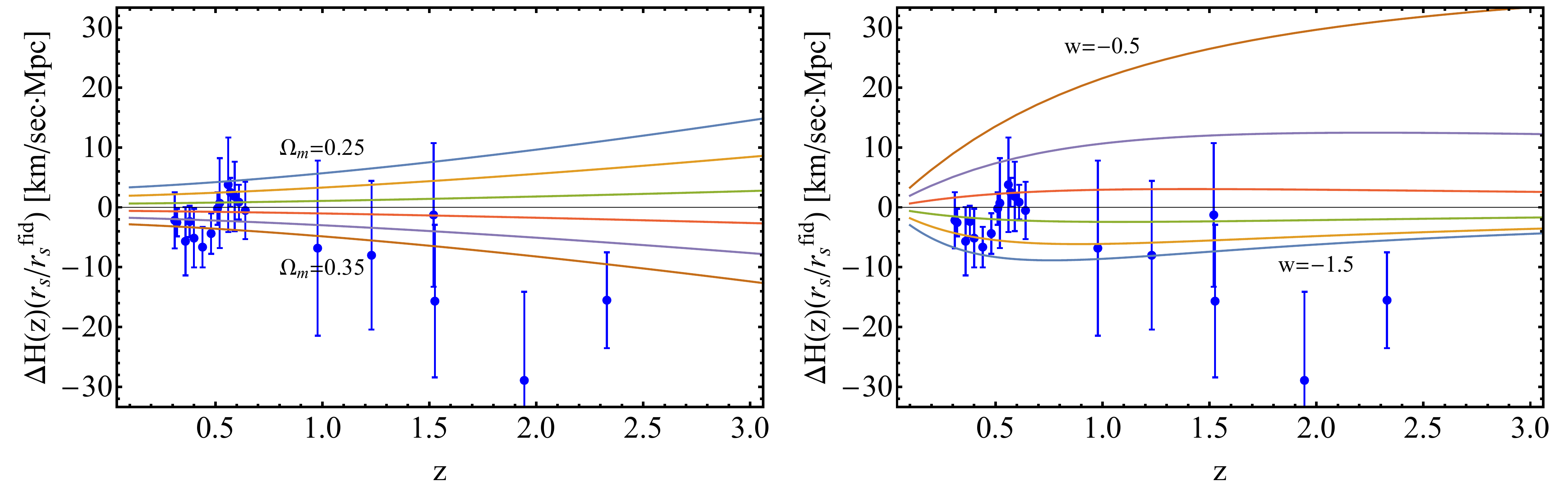}
\par\end{centering}
\caption{The deviation $\Delta$\baohspace as a function of the redshift $z$ for different values of \omm (left panel) and $w$ (right panel)}  
\label{fig:hfig}
\end{figure*}

\begin{figure*} 
\begin{centering}
\includegraphics[width =0.95\textwidth]{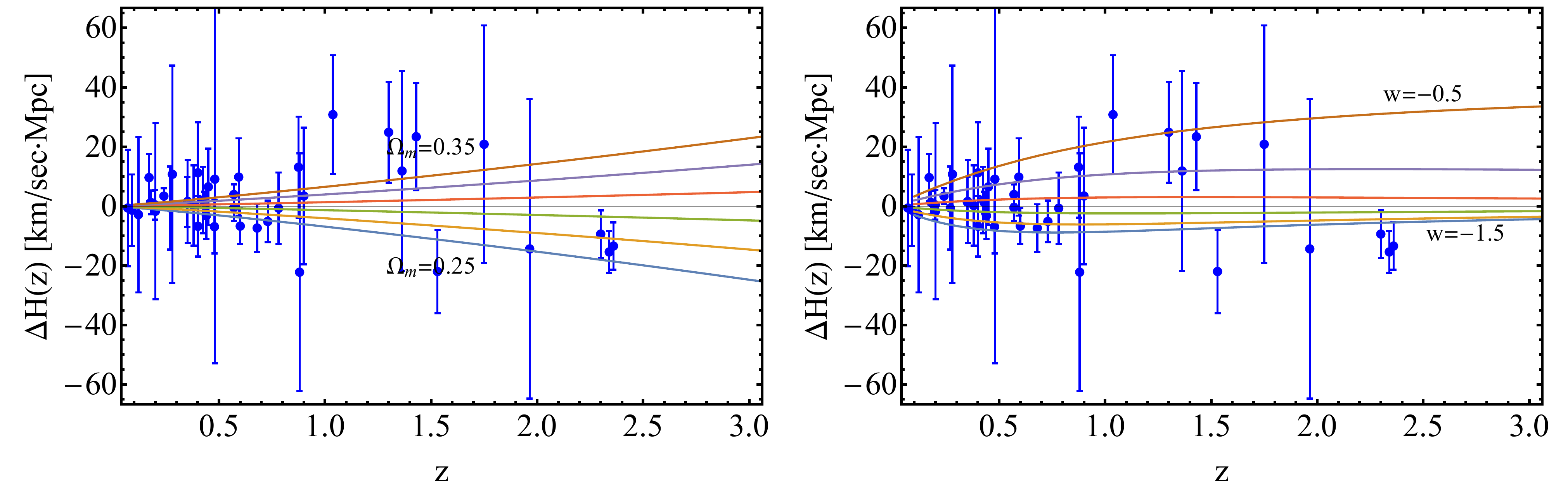}
\par\end{centering}
\caption{The deviation $\Delta H(z)$ as a function of redshift using the full compilation of Table \ref{tab:data-hz} in the Appendix, for various values of \omm (left panel) and $w$ (right panel).}  
\label{fig:hdatfig}
\end{figure*}
\end{widetext}  

In Fig. \ref{fig:dvfig}  we show the predicted evolution of the deviation of the observable \baodvspace for various values of \omm (left panel) and of $w$ (right panel). The deviation of the parameter \omm (left panel) was performed around the value $\Omega_{m}=0.3$ while the deviation of the parameter $w$ was performed around the \lcdm value $w=-1$ [see Eq. \eqref{eq:devdef}]. Notice the existence of a blind spot at $z\simeq 1.2$ for the observable \baodvspace with respect to the parameter \ommnospace, while the optimal redshift in the same plot is $z\simeq 0.6$. (Even though the region $z>2$ also seems to be optimal, there are currently almost no data available in this redshift range). In contrast, for the same observable with respect to the parameter $w$ there is no blind spot, while the optimal redshift range is at $z> 1.2$.
 
\begin{figure*}
\begin{centering}
\includegraphics[width =0.95\textwidth]{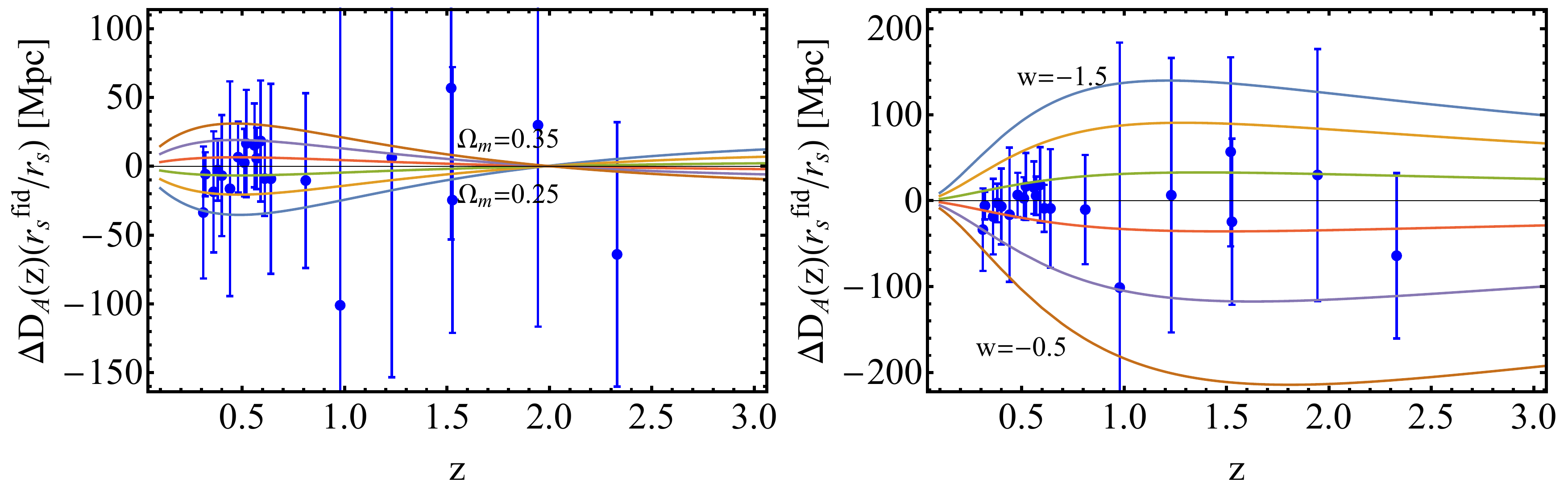}
\par\end{centering}
\caption{The deviation $\Delta$\baodaspace as a function of the redshift $z$ for different values of \omm (left panel) and $w$ (right panel)}  
\label{fig:dafig}
\end{figure*}
 
In Fig. \ref{fig:hfig} we show the predicted evolution of the deviation of the observable \baohspace for various values of \omm (left panel) and of $w$ (right panel). For this observable there is no blind redshift spot, while the sensitivity appears to increase monotonically with redshift for both observables. Notice the asymmetry obtained for the equation of state parameter which is due to the fact that for $w<-1$ at early times the effects of dark energy are negligible for all values of $w$, leading to a degeneracy for this range of parameters at high $z$. For comparison, in Fig. \ref{fig:hdatfig}, we show the deviation of the observable Hubble expansion rate for various values of \omm (left panel) and of $w$ (right panel) along with corresponding data obtained from the spectroscopic evolution of galaxies used as cosmic chronometers, shown in Table \ref{tab:data-hz} in the Appendix along with the corresponding citations (for previous compilations see also Refs. \cite{Moresco:2016mzx,Anagnostopoulos:2017iao,Guo:2015gpa}). Even though Figs. \ref{fig:hfig} and \ref{fig:hdatfig} are qualitatively similar, it is clear that the BAO data are significantly more constraining compared to the cosmic chronometer data with respect to both parameters \omm and $w$, especially at low redshifts. 

In Fig. \ref{fig:dafig} we show the predicted evolution of the deviation of the observable \baodaspace for various values of \omm (left panel) and  $w$ (right panel). The behavior of this observable is similar to that of \baodvspace even though the blind spot with respect to the parameter \omm appears at a higher redshift ($z\simeq 2$), while at higher redshifts the sensitivity of this observable with respect to the parameter \omm is significantly reduced compared to the sensitivity of \baodv.

A comparison of the three BAO observable distances $\frac{D_M(z)}{r_s\sqrt{z}}$, $\frac{D_V(z)}{r_s \sqrt{z}}$ and $\frac{zD_H(z)}{r_s \sqrt{z}}$ [as $D_H(z)=\frac{c}{H(z)}$]  for the \plcdm best fit parameter values  along with the corresponding data from Table \ref{tab:data-bao} of the Appendix is shown in Fig. \ref{fig:BAOdist}.
\begin{figure}
\begin{centering}
\includegraphics[width =0.45\textwidth]{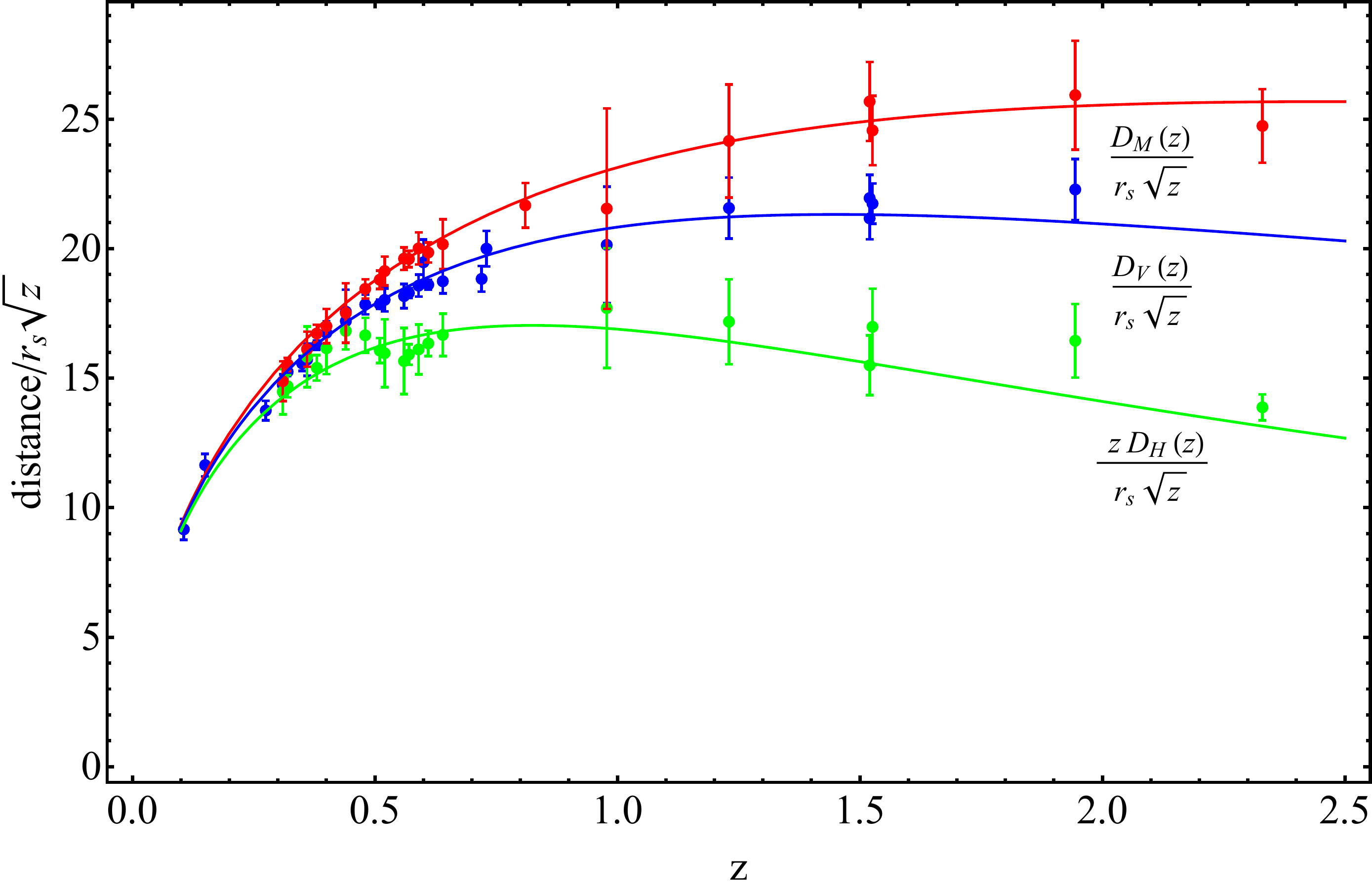}
\par\end{centering}
\caption{The BAO observable distances for the \plcdm best-fit parameter values  along with the corresponding data from Table \ref{tab:data-bao} in the Appendix. The data appear to be in good agreement with the \plcdm predictions.}   
\label{fig:BAOdist}
\end{figure} 
This plot is in excellent agreement with the corresponding plot of Ref. \cite{Alam:2016hwk}  (Fig. 14) even though here we superpose the \plcdm prediction with a significantly larger compilation of datapoints. As demonstrated in the next subsection the BAO data are in good agreement with the \plcdm parameter values.

\subsection{Contour Shapes and Redshift Ranges}
The existence of optimal and blind redshift ranges for the BAO observables with respect to cosmological parameters has an effect on the form of maximum likelihood contours obtained from data at various redshift ranges. In particular, the  Figure of Merit (reciprocal of the area of confidence contours in parameter space) tends to decrease for datasets  with redshifts close to blind redshift spots and increase for datasets with redshifts close to optimal redshift regions. In order to demonstrate this effect, we construct the confidence contours for the parameters \omm and $w$ using the BAO observables in different redshift regions.

In order to construct $\chi^2$ we first consider the vector
\be 
V^i_{BAO}(z_i,\Omega_{m},w)\equiv BAO^m_{i} - BAO^m_{theoretical}\label{eq:vechi}
\ee
\begin{widetext}
\begin{figure*}
\begin{centering}
\includegraphics[width =0.95\textwidth]{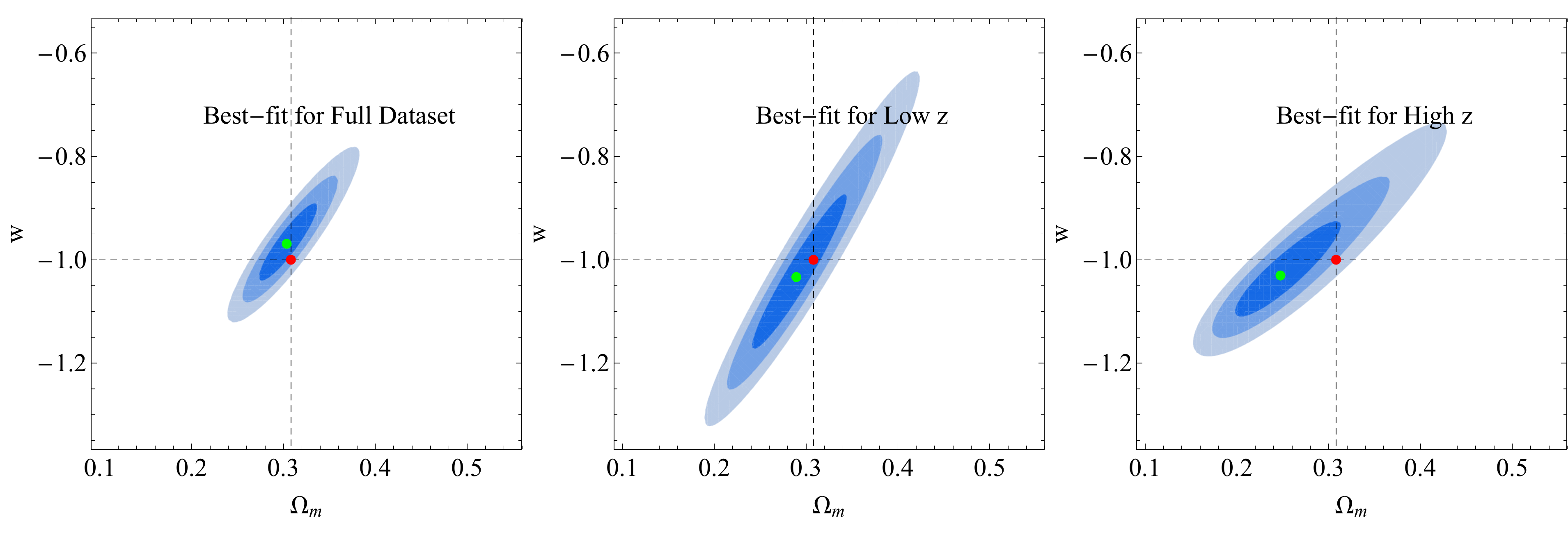}
\par\end{centering}
\caption{The $1\sigma-3\sigma$ contours in the $\Omega_m-w$ parametric space. The contours describe the corresponding confidence regions using the full compilation of \baodvspace data (left panel), low redshift ($z<0.55$) data (middle panel) and high redshift ($z>0.55$) data (right panel) from Table \ref{tab:data-bao} in the Appendix. The red and green dots describe the \plcdm best fit and the best-fit values from the compilation of \baodvspace  data. Notice that at high $z$ close to the blind spot for $\Omega_{m}$ and the optimum redshift for $w$, the thickness of the contours (uncertainty)  increases along the $\Omega_{m}$ axis and decreases along the $w$ axis (the contours are rotated clockwise) as expected from Fig. \ref{fig:dvfig}. }  
\label{fig:contourdvfig}
\end{figure*}
\end{widetext} 
where $m$ runs from 1 to 3 indicating the different types of BAO data of Table \ref{tab:data-bao} in the Appendix and the theoretical expressions for \baoda, \baodvspace and \baohspace are given in Eqs. \eqref{dazdef}, \eqref{dvdef} and \eqref{eq:hubble} respectively. $\chi^2$ is obtained as
\be
\chi^2 = V^i F_{ij} V^j 
\label{eq:chi2}
\ee
where $F_{ij}$ is the Fisher matrix (inverse of the covariance matrix $C_{ij}$). 

The covariance matrix for the \baodvspace  data takes the form 
\be
C_{ij,D_V \times \left(r_s^{fid.}/r_s \right)}^{\textrm{BAO,total}}=\left(
         \begin{array}{cccc}
           \sigma_1^2 & 0 & 0 & \cdots \\
           0 & C_{ij}^{WiggleZ} & 0& \cdots \\
           0 & 0 & \cdots &   \sigma_N^2 \\
         \end{array}
       \right) \label{eq:totalcij}
\ee
where $N=28$ and \cite{Kazin:2014qga}
\be
C_{ij}^{WiggleZ}=F_{ij,\text{WiggleZ}}^{-1}=10^{4}\left(
         \begin{array}{ccc}
           2.18 & -1.12 & 0.47 \\
          -1.12 & 1.71 & -0.72 \\
           0.47 & -0.72 & 1.65 \\
         \end{array}
       \right)^{-1} \label{eq:wigglez}
\ee
whereas for both \baodaspace and \baohspace we have assumed a diagonal covariance matrix
\be
C_{ij}^{\textrm{BAO,total}}=\left(
         \begin{array}{cccc}
           \sigma_1^2 & 0 & 0 & \cdots \\
           0 & \sigma_2^2 & 0& \cdots \\
           0 & 0 & \cdots &   \sigma_N^2 \\
         \end{array}
       \right) \label{eq:totalcijother}
\ee
where $N$ is equal to the considered number of datapoints. 

The forms of Eqs. \eqref{eq:totalcij} and \eqref{eq:totalcijother} are clearly oversimplifications of the actual covariance matrices, since these forms ignore possible correlations between the considered BAO data. However, to the best of our knowledge the non-diagonal terms of the $D_A$ and $H$ covariance matrices are not publicly available. In order to estimate the magnitude of the effects of these terms we have performed Monte Carlo simulations including random nondiagonal terms to the covariance matrices for $D_A$ and $H$ of relative magnitude similar to the nondiagonal terms of the nondiagonal terms corresponding to $D_V$ setting the magnitude of the matrix \cite{Kazantzidis:2018rnb}
\be 
C_{ij}=\frac{1}{2} \sigma_i \cdot \sigma_j
\ee
where $\sigma_i$ and $\sigma_j$ are the errors of the published datapoints $i$ and $j$ respectively. These simulations indicated that the likelihood  contours and the best fit parameter values do not change more than $10\%$ when we include the nondiagonal terms in the covariance matrix. Thus, possible reasonable correlations among datapoints are not expected to significantly affect our results \cite{suppl}. 

In the left panel of Fig. \ref{fig:contourdvfig} we show the $1\sigma-3\sigma$ $\Omega_{m}-w$ contour plots for the full \baodvspace data of Table \ref{tab:data-bao} in the Appendix using Eqs. \eqref{eq:vechi}-\eqref{eq:totalcij} and ignoring the possible correlations among the datapoints. The best fit parameter values are within $1\sigma$ from the corresponding best fit \plcdm values (red dot). 

Furthermore we construct the same contour plots for low-redshift \baodvspace data (middle panel of Fig. \ref{fig:contourdvfig}), where $z<0.55$ (14 datapoints), and for high-redshift  \baodvspace data (right panel of Fig. \ref{fig:contourdvfig}), where $z>0.55$ (14 datapoints). The low-redshift data correspond to optimal redshift for the parameter \omm (see Fig. \ref{fig:dvfig}) and thus the confidence contours are thinner in the direction of the \omm axis while the contours are elongated in the $w$ direction. In contrast the high-redshift data are close  to the \omm blind spot and thus the confidence contours are thicker in the \omm direction (left panel), while the contours are suppressed in the $w$ direction (as expected from Fig. \ref{fig:dvfig}) which indicates an optimal high-redshift range for the parameter $w$.

Similar conclusions and confidence contours  are obtained from the low and high redshift data for \baodaspace and \baohspace data (see Supplemental Material \cite{suppl}).
\begin{widetext}
\begin{figure*}
\begin{centering}
\includegraphics[width =0.98\textwidth]{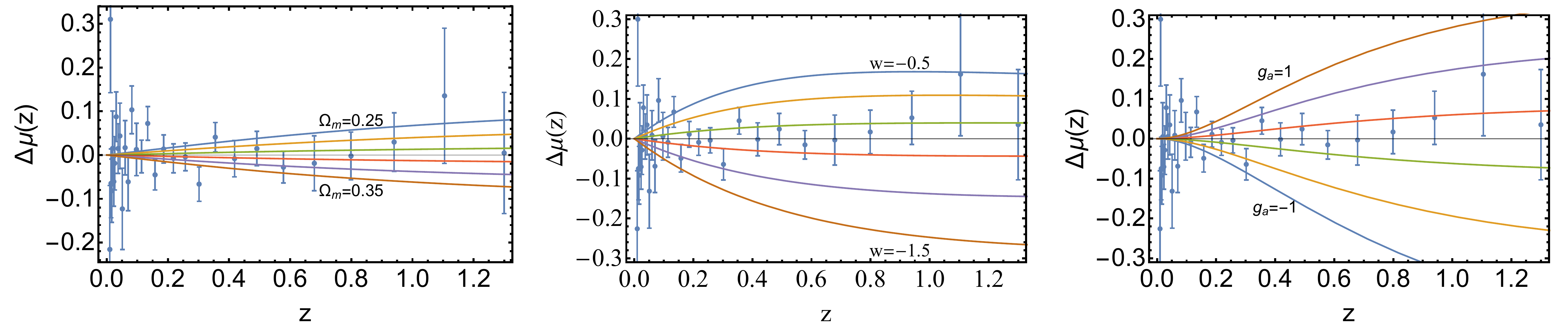}
\par\end{centering}
\caption{The deviation of the distance modulus observable $\Delta \mu$ as a function of redshift  for \omm (left panel), $w$ (middle panel) and $g_\alpha$ (right panel) superimposed with the JLA data of Table \ref{tab:data-jla} in the Appendix.}  
\label{fig:mufig}
\end{figure*}
\begin{figure*} 
\begin{centering}
\includegraphics[width =0.98\textwidth]{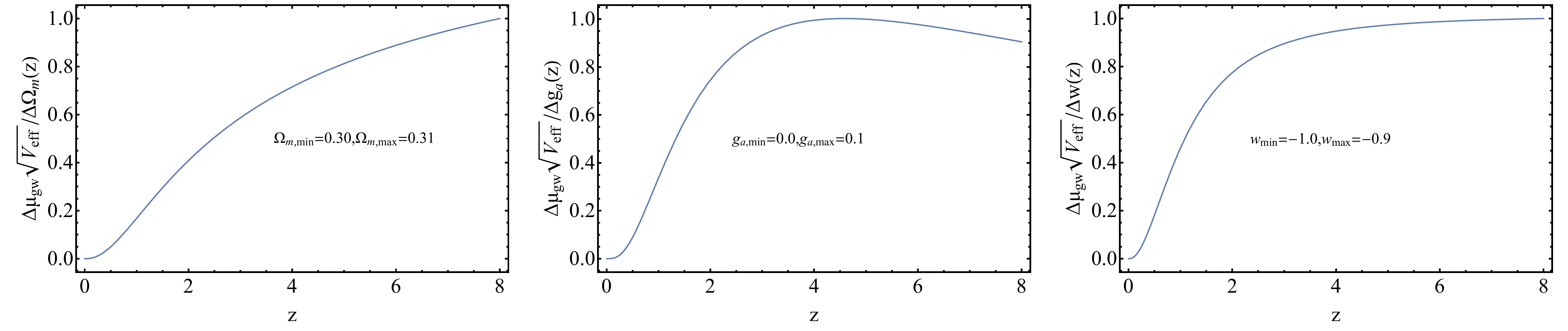}
\par\end{centering}
\caption{The sensitivity measure as a function of redshift  $z$  for \omm (left panel), $g_\alpha$ (middle panel) and $w$ (right panel).}   
\label{fig:sensmugwsens} 
\end{figure*}
\end{widetext}  
\section{Distance Moduli from SnIa and from Gravitational Waves}
\label{sec:Distance Moduli}
The luminosity distance 
\be 
D_L(z;\Omega_{m},w)=(1+z)\int_0^z\frac{c\; dz}{H(z;\Omega_{m},w)}
\label{dlzdefdist}
\ee
is an important cosmological observable that is measured using standard candles like SnIa or standard gravitational wave sirens, like merging binary neutron star systems observed via multi-messenger observations.
The distance modulus $\mu =m-M$ is the difference between the apparent magnitude  $m$ and the absolute magnitude M of standard candle. It is related to the luminosity distance $D_L$ in Mpc as
\be
\mu(z;\Omega_{m},w)=5 log_{10}(D_L)+25  
\label{mudef1}
\ee
In the context of a varying effective Newton's constant $G_{eff}(z)$ the absolute magnitude of SnIa is expected to vary with redshift as \cite{GarciaBerro:1999bq,Gaztanaga:2001fh,Nesseris:2006jc}       
\be 
M-M_0=\frac{15}{4}log_{10} \left(\frac{G_{eff}}{G_N} \right)  
\label{absmagvar}
\ee
where the subscript $0$ refers to local value of $M$. Thus, for SnIa $\mu$ also depends on the evolution of $G_{eff}(z)$ (or equivalently on the parameter $g_a$) as
\be
\mu(z;\Omega_{m},w,g_a)=5 log_{10}(D_L)+\frac{15}{4}log_{10}\left(\frac{G_{eff}(z;g_a)}{G_N} \right)+25
\label{mudef2}
\ee 
In the case of gravitational wave luminosity distance,  the corresponding gravitational wave distance modulus obtained from standard sirens is of the form \cite{Belgacem:2017ihm} 
\be
\mu_{gw}(z;\Omega_{m},w,g_a)=5 log_{10} \left(D_L \sqrt{\frac{G_{eff}}{G_N}} \right)+25
\label{mudefgw}
\ee
In Fig. \ref{fig:mufig} we show the deviation $\Delta \mu$ as a function of redshift  for \omm (left panel), $w$ (middle panel) and $g_\alpha$ (right panel) superimposed with the JLA SnIa binned data of Table \ref{tab:data-jla} in the Appendix. The corresponding sensitivity measure is shown in Fig. \ref{fig:sensmugwsens}. Notice that even though the deviation $\Delta \mu_{gw}$ appears to be increasing with redshift for all the parameters considered, the absolute value of the sensitivity measure with respect to the parameter $g_a$ has a maximum for redshifts in the range  $z\in [4,5]$, indicating the presence of an optimal redshift range.  

\begin{figure}
\begin{centering}
\includegraphics[width =0.45\textwidth]{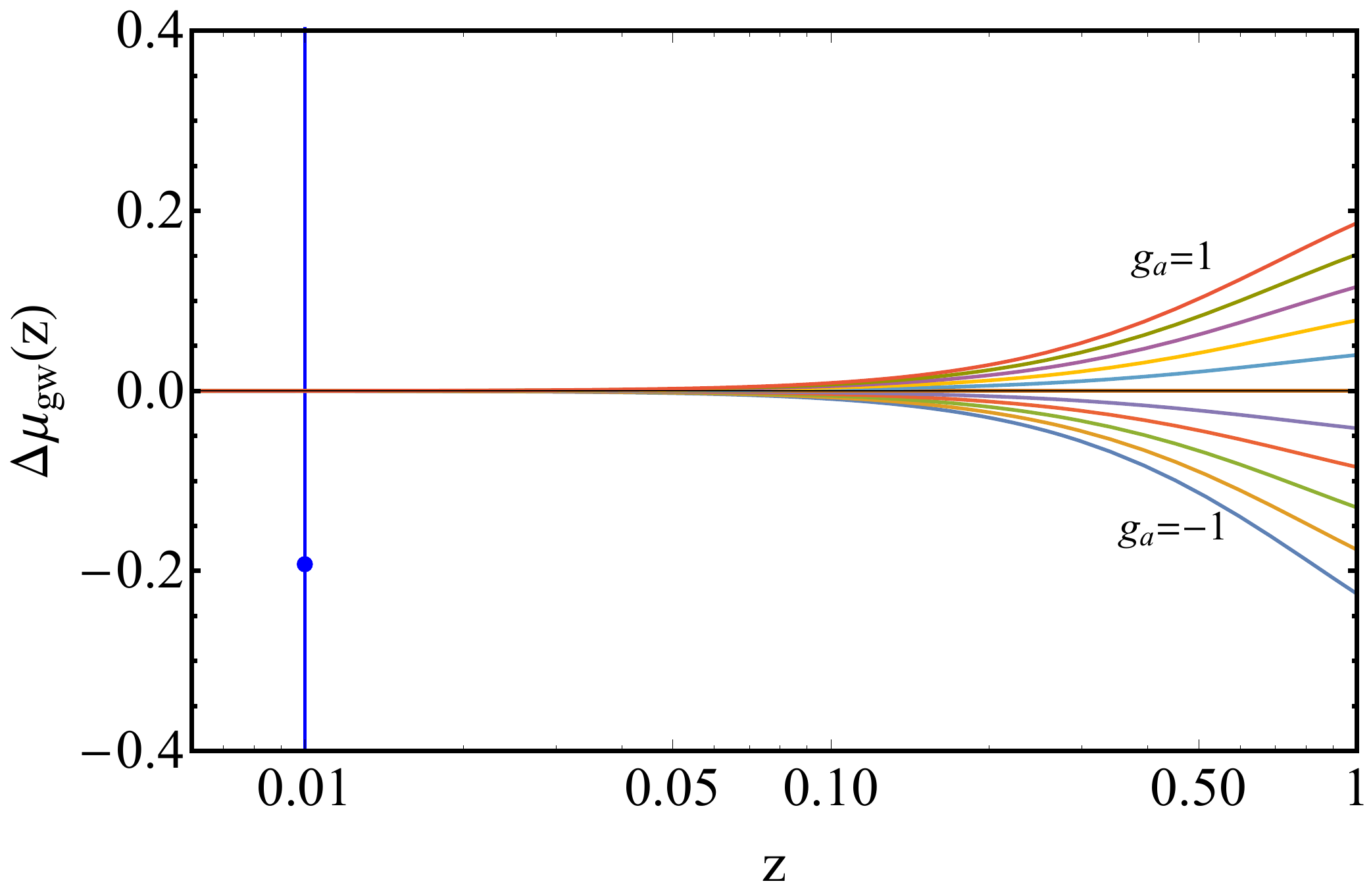}
\par\end{centering}
\caption{The deviation of the gravitational wave distance modulus with the parameter $g_a$. The only existing datapoint does not lead to any useful constraints.}  
\label{fig:gwga}
\end{figure}

The deviations  $\Delta\mu_{gw}(z)$ with respect to the parameters \omm and $w$ is identical to the corresponding deviations $\Delta \mu(z)$, since for $g_a=0$ we have $\Delta \mu(z)=\Delta \mu_{gw}(z)$. The deviation $\Delta\mu_{gw}(z)$ with respect to the parameter $g_a$ is shown in Fig. \ref {fig:gwga} 
along with the single available datapoint from the standard siren GW170817 \cite{PhysRevLett.119.161101,Abbott:2017xzu}. Clearly even though standard siren data can in principle be used to constrain the evolution of $G_{eff}$, a dramatic improvement is required before such probes become competitive with growth and SnIa data.

\section{Discussion-Outlook}
\label{sec:Discussion} 

We have demonstrated that the constraining power (sensitivity) of a wide range of cosmological observables on cosmological parameters is a rapidly varying function of the redshift where the observable is measured. In fact, this sensitivity in many cases does not vary monotonically with redshift but has degeneracy points (redshift blind spots) and maxima (optimal redshift ranges) which are relatively close in redshift space. The identification of such regions can contribute to the optimal design and redshift range selection of cosmological probes aimed at constraining specific cosmological parameters through measurement of cosmological observables. In addition, we have shown that many of the recent \fs RSD data, which tend to be at higher redshifts ($z>0.8$) are close to blind spots of the observable \fs with respect to all three cosmological parameters considered (\ommnospace, $w$ and $g_a$). A similar trend for probing higher redshifts also exists for upcoming surveys as demonstrated in Table \ref{table:surveys}. A more efficient strategy for this observable would be an improvement of the measurements at lower redshifts instead of focusing on higher redshifts. Such a strategy would lead to improved constraints on all three parameters considered.

Even though our analysis has revealed the generic existence of optimal redshifts and blind spots of observables with respect to specific cosmological parameters, it still has not taken into account all relevant effects that play a role in determining the exact location of these points in redshift space. For example, we have not explicitly taken into account the number of linear modes available to a survey in redshift space as well as the dependence of the effective volume $V_{eff}$ on the number of tracers and their selection. We anticipate that these effects could mildly shift the location of the derived blind spots and optimal redshifts determined by our analysis.

An interesting extension of our analysis could involve the consideration of other observables and additional cosmological parameters (e.g. an equation of state parameter that evolves with redshift). The existence of blind spots could be avoided by considering various functions and/or combinations of cosmic observables designed in such a way as to optimize sensitivity for given cosmological parameters in a given redshift range. The investigation of the efficiency of such combinations is also an interesting extension of this project.

\section*{Acknowledgements}
We thank the anonymous referee for insightful comments that improved the quality of our paper. This research is co-financed by Greece and the European Union (European Social Fund- ESF) through the Operational Programme ``Human Resources Development, Education and Lifelong Learning" in the context of the project ``Strengthening Human Resources Research Potential via Doctorate Research" (MIS-5000432), implemented by the State Scholarships Foundation (IKY). This article has benefited from COST Action CA15117 (CANTATA), supported by COST (European Cooperation in Science and Technology).

\textbf{Supplemental Material:} The Mathematica files used for the numerical analysis and for construction of the figures can be found in \cite{suppl}.

\appendix
\section{Data Used in the Analysis}
\label{sec:Appendix_A}
In this appendix we present the data used in the analysis. 

\begin{widetext}
\begin{longtable}{ | c | c | c | c | c | c | c | }
\caption{The compilation of RSD data used in the present analysis and in the analysis of Ref. \cite{Kazantzidis:2018rnb}} 
\label{tab:data-rsd}\\
\hline
    Index & Dataset & $z$ & $f\sigma_8(z)$ & Refs. & Year & Fiducial Cosmology \\
\hline
1 & SDSS-LRG & $0.35$ & $0.440\pm 0.050$ & \cite{Song:2008qt} &  30 October 2006 &$(\Omega_{m},\Omega_K,\sigma_8$)$=(0.25,0,0.756)$\cite{Tegmark:2006az} \\

2 & VVDS & $0.77$ & $0.490\pm 0.18$ & \cite{Song:2008qt}  & 6 October 2009 & $(\Omega_{m},\Omega_K,\sigma_8)=(0.25,0,0.78)$ \\

3 & 2dFGRS & $0.17$ & $0.510\pm 0.060$ & \cite{Song:2008qt}  &  6 October 2009 & $(\Omega_{m},\Omega_K)=(0.3,0,0.9)$ \\

4 & 2MRS &0.02& $0.314 \pm 0.048$ &  \cite{Davis:2010sw}, \cite{Hudson:2012gt}& 13 Novemver 2010 & $(\Omega_{m},\Omega_K,\sigma_8)=(0.266,0,0.65)$ \\

5 & SnIa+IRAS &0.02& $0.398 \pm 0.065$ & \cite{Turnbull:2011ty}, \cite{Hudson:2012gt}& 20 October 2011 & $(\Omega_{m},\Omega_K,\sigma_8)=(0.3,0,0.814)$\\

6 & SDSS-LRG-200 & $0.25$ & $0.3512\pm 0.0583$ & \cite{Samushia:2011cs} & 9 December 2011 & $(\Omega_{m},\Omega_K,\sigma_8)=(0.276,0,0.8)$  \\

7 & SDSS-LRG-200 & $0.37$ & $0.4602\pm 0.0378$ & \cite{Samushia:2011cs} & 9 December 2011 & \\

8 & SDSS-LRG-60 & $0.25$ & $0.3665\pm0.0601$ & \cite{Samushia:2011cs} & 9 December 2011 & $(\Omega_{m},\Omega_K,\sigma_8)=(0.276,0,0.8)$ \\

9 & SDSS-LRG-60 & $0.37$ & $0.4031\pm0.0586$ & \cite{Samushia:2011cs} & 9 December 2011 &\\

10 & WiggleZ & $0.44$ & $0.413\pm 0.080$ & \cite{Blake:2012pj} & 12 June 2012  & $(\Omega_{m},h,\sigma_8)=(0.27,0.71,0.8)$ \\

11 & WiggleZ & $0.60$ & $0.390\pm 0.063$ & \cite{Blake:2012pj} & 12 June 2012 &  \\

12 & WiggleZ & $0.73$ & $0.437\pm 0.072$ & \cite{Blake:2012pj} & 12 June 2012 &\\

13 & 6dFGS& $0.067$ & $0.423\pm 0.055$ & \cite{Beutler:2012px} & 4 July 2012 & $(\Omega_{m},\Omega_K,\sigma_8)=(0.27,0,0.76)$ \\

14 & SDSS-BOSS& $0.30$ & $0.407\pm 0.055$ & \cite{Tojeiro:2012rp} & 11 August 2012 & $(\Omega_{m},\Omega_K,\sigma_8)=(0.25,0,0.804)$ \\

15 & SDSS-BOSS& $0.40$ & $0.419\pm 0.041$ & \cite{Tojeiro:2012rp} & 11 August 2012 & \\

16 & SDSS-BOSS& $0.50$ & $0.427\pm 0.043$ & \cite{Tojeiro:2012rp} & 11 August 2012 & \\

17 & SDSS-BOSS& $0.60$ & $0.433\pm 0.067$ & \cite{Tojeiro:2012rp} & 11 August 2012 & \\

18 & Vipers& $0.80$ & $0.470\pm 0.080$ & \cite{delaTorre:2013rpa} & 9 July 2013 & $(\Omega_{m},\Omega_K,\sigma_8)=(0.25,0,0.82)$  \\

19 & SDSS-DR7-LRG & $0.35$ & $0.429\pm 0.089$ & \cite{Chuang:2012qt}  & 8 August 2013 & $(\Omega_{m},\Omega_K,\sigma_8$)$=(0.25,0,0.809)$\cite{Komatsu:2010fb}\\

20 & GAMA & $0.18$ & $0.360\pm 0.090$ & \cite{Blake:2013nif}  & 22 September 2013 & $(\Omega_{m},\Omega_K,\sigma_8)=(0.27,0,0.8)$ \\

21& GAMA & $0.38$ & $0.440\pm 0.060$ & \cite{Blake:2013nif}  & 22 September 2013 & \\

22 & BOSS-LOWZ& $0.32$ & $0.384\pm 0.095$ & \cite{Sanchez:2013tga}  & 17 December 2013  & $(\Omega_{m},\Omega_K,\sigma_8)=(0.274,0,0.8)$ \\

23 & SDSS DR10 and DR11 & $0.32$ & $0.48 \pm 0.10$ & \cite{Sanchez:2013tga} &   17 December 2013 & $(\Omega_{m},\Omega_K,\sigma_8$)$=(0.274,0,0.8)$\cite{Anderson:2013zyy}\\

24 & SDSS DR10 and DR11 & $0.57$ & $0.417 \pm 0.045$ & \cite{Sanchez:2013tga} &  17 December 2013 &  \\

25 & SDSS-MGS & $0.15$ & $0.490\pm0.145$ & \cite{Howlett:2014opa} & 30 January 2015 & $(\Omega_{m},h,\sigma_8)=(0.31,0.67,0.83)$ \\

26 & SDSS-veloc & $0.10$ & $0.370\pm 0.130$ & \cite{Feix:2015dla}  & 16 June 2015 & $(\Omega_{m},\Omega_K,\sigma_8$)$=(0.3,0,0.89)$\cite{Tegmark:2003uf} \\

27 & FastSound& $1.40$ & $0.482\pm 0.116$ & \cite{Okumura:2015lvp}  & 25 November 2015 & $(\Omega_{m},\Omega_K,\sigma_8$)$=(0.27,0,0.82)$\cite{Hinshaw:2012aka} \\

28 & SDSS-CMASS & $0.59$ & $0.488\pm 0.060$ & \cite{Chuang:2013wga} & 8 July 2016 & $\ \ (\Omega_{m},h,\sigma_8)=(0.307115,0.6777,0.8288)$ \\

29 & BOSS DR12 & $0.38$ & $0.497\pm 0.045$ & \cite{Alam:2016hwk} & 11 July 2016 & $(\Omega_{m},\Omega_K,\sigma_8)=(0.31,0,0.8)$ \\

30 & BOSS DR12 & $0.51$ & $0.458\pm 0.038$ & \cite{Alam:2016hwk} & 11 July 2016 & \\

31 & BOSS DR12 & $0.61$ & $0.436\pm 0.034$ & \cite{Alam:2016hwk} & 11 July 2016 & \\

32 & BOSS DR12 & $0.38$ & $0.477 \pm 0.051$ & \cite{Beutler:2016arn} & 11 July 2016 & $(\Omega_{m},h,\sigma_8)=(0.31,0.676,0.8)$ \\

33 & BOSS DR12 & $0.51$ & $0.453 \pm 0.050$ & \cite{Beutler:2016arn} & 11 July 2016 & \\

34 & BOSS DR12 & $0.61$ & $0.410 \pm 0.044$ & \cite{Beutler:2016arn} & 11 July 2016 &  \\

35 &Vipers v7& $0.76$ & $0.440\pm 0.040$ & \cite{Wilson:2016ggz} & 26 October 2016  & $(\Omega_{m},\sigma_8)=(0.308,0.8149)$ \\

36 &Vipers v7 & $1.05$ & $0.280\pm 0.080$ & \cite{Wilson:2016ggz} & 26 October 2016 &\\

37 &  BOSS LOWZ & $0.32$ & $0.427\pm 0.056$ & \cite{Gil-Marin:2016wya} & 26 October 2016 & $(\Omega_{m},\Omega_K,\sigma_8)=(0.31,0,0.8475)$\\

38 & BOSS CMASS & $0.57$ & $0.426\pm 0.029$ & \cite{Gil-Marin:2016wya} & 26 October 2016 & \\

39 & Vipers  & $0.727$ & $0.296 \pm 0.0765$ & \cite{Hawken:2016qcy} &  21 November 2016 & $(\Omega_{m},\Omega_K,\sigma_8)=(0.31,0,0.7)$\\

40 & 6dFGS+SnIa & $0.02$ & $0.428\pm 0.0465$ & \cite{Huterer:2016uyq} & 29 November 2016 & $(\Omega_{m},h,\sigma_8)=(0.3,0.683,0.8)$ \\

41 & Vipers  & $0.6$ & $0.48 \pm 0.12$ & \cite{delaTorre:2016rxm} & 16 December 2016 & $(\Omega_{m},\Omega_b,n_s,\sigma_8$)= $(0.3, 0.045, 0.96,0.831)$\cite{Ade:2015xua} \\

42 & Vipers  & $0.86$ & $0.48 \pm 0.10$ & \cite{delaTorre:2016rxm} & 16 December 2016  & \\

43 &Vipers PDR-2& $0.60$ & $0.550\pm 0.120$ & \cite{Pezzotta:2016gbo} & 16 December 2016 & $(\Omega_{m},\Omega_b,\sigma_8)=(0.3,0.045,0.823)$ \\

44 & Vipers PDR-2& $0.86$ & $0.400\pm 0.110$ & \cite{Pezzotta:2016gbo} & 16 December 2016 &\\

45 & SDSS DR13  & $0.1$ & $0.48 \pm 0.16$ & \cite{Feix:2016qhh} & 22 December 2016 & $(\Omega_{m},\sigma_8$)$=(0.25,0.89)$\cite{Tegmark:2003uf} \\

46 & 2MTF & 0.001 & $0.505 \pm 0.085$ &  \cite{Howlett:2017asq} & 16 June 2017 & $(\Omega_{m},\sigma_8)=(0.3121,0.815)$\\

47 & Vipers PDR-2 & $0.85$ & $0.45 \pm 0.11$ & \cite{Mohammad:2017lzz} & 31 July 2017  &  $(\Omega_b,\Omega_{m},h)=(0.045,0.30,0.8)$ \\

48 & BOSS DR12 & $0.31$ & $0.469 \pm 0.098$ &  \cite{Wang:2017wia} & 15 September 2017 & $(\Omega_{m},h,\sigma_8)=(0.307,0.6777,0.8288)$\\

49 & BOSS DR12 & $0.36$ & $0.474 \pm 0.097$ &  \cite{Wang:2017wia} & 15 September 2017 & \\

50 & BOSS DR12 & $0.40$ & $0.473 \pm 0.086$ &  \cite{Wang:2017wia} & 15 September 2017 & \\

51 & BOSS DR12 & $0.44$ & $0.481 \pm 0.076$ &  \cite{Wang:2017wia} & 15 September 2017 & \\

52 & BOSS DR12 & $0.48$ & $0.482 \pm 0.067$ &  \cite{Wang:2017wia} & 15 September 2017 & \\

53 & BOSS DR12 & $0.52$ & $0.488 \pm 0.065$ &  \cite{Wang:2017wia} & 15 September 2017 & \\

54 & BOSS DR12 & $0.56$ & $0.482 \pm 0.067$ &  \cite{Wang:2017wia} & 15 September 2017 & \\

55 & BOSS DR12 & $0.59$ & $0.481 \pm 0.066$ &  \cite{Wang:2017wia} & 15 September 2017 & \\

56 & BOSS DR12 & $0.64$ & $0.486 \pm 0.070$ &  \cite{Wang:2017wia} & 15 September 2017 & \\

57 & SDSS DR7 & $0.1$ & $0.376\pm 0.038$ & \cite{Shi:2017qpr} & 12 December 2017 & $(\Omega_{m},\Omega_b,\sigma_8)=(0.282,0.046,0.817)$ \\

58 & SDSS-IV & $1.52$ & $0.420 \pm 0.076$ &  \cite{Gil-Marin:2018cgo} & 8 January 2018  & $(\Omega_{m},\Omega_b h^2,\sigma_8)=(0.26479, 0.02258,0.8)$ \\ 

59 & SDSS-IV & $1.52$ & $0.396 \pm 0.079$ & \cite{Hou:2018yny} & 8 January 2018 & $(\Omega_{m},\Omega_b h^2,\sigma_8)=(0.31,0.022,0.8225)$ \\ 

60 & SDSS-IV & $0.978$ & $0.379 \pm 0.176$ &  \cite{Zhao:2018jxv} & 9 January 2018 &$(\Omega_{m},\sigma_8)=(0.31,0.8)$\\

61 & SDSS-IV & $1.23$ & $0.385 \pm 0.099$ &  \cite{Zhao:2018jxv} & 9 January 2018 & \\

62 & SDSS-IV & $1.526$ & $0.342 \pm 0.070$ &  \cite{Zhao:2018jxv} & 9 January 2018 & \\

63 & SDSS-IV & $1.944$ & $0.364 \pm 0.106$ &  \cite{Zhao:2018jxv} & 9 January 2018 & \\
\hline
\end{longtable}

\begin{longtable}{ | c | c | c | c | c | c | c | }
\caption{A compilation of BAO data that have been published from 2006 until 2018 in chronological order}
\label{tab:data-bao}\\
\hline
   Index & $z_{eff}$ & $D_A \times \left(r_s^{fid.}/r_s \right)$ (Mpc) & $H(z) \times \left(r_s/r_s^{fid.} \right)$ (km/sec Mpc) & $D_V \times \left(r_s^{fid.}/r_s \right)$ (Mpc) & Year & Ref.  \\
\hline
1 & $0.275$ & - & - & $1061.87 \pm 29$  &  2 November 2009  & \cite{Percival:2009xn} \\
2 & $0.106$ & - & - & $439.3 \pm 19.6$ &  16 June 2011 & \cite{Beutler:2011hx} \\
3 & $0.35$ & - & - & $1356 \pm 25$ &  28 March 2012 & \cite{Mehta:2012hh} \\
4 & $0.44$ & - & - & $1716 \pm 83$ &  28 July 2014 & \cite{Kazin:2014qga} \\
5 & $0.60$ & - & - & $2221 \pm 100$ &  28 July 2014 & \cite{Kazin:2014qga} \\
6 & $0.73$ & - & - & $2516 \pm 86$ &  28 July 2014  & \cite{Kazin:2014qga} \\
7 & $0.15$ & - & - & $664 \pm 25$ &  21 January 2015 & \cite{Ross:2014qpa} \\
8 & $0.38$ & $1100 \pm 22$ & $81.5 \pm 2.6$ & $1477 \pm 16$ & 11 July 2016 & \cite{Alam:2016hwk} \\
9 & $0.51$ & $1309.3 \pm 24.5$ & $90.5 \pm 2.7$ & $1877 \pm 19$ &  11 July 2016 & \cite{Alam:2016hwk} \\
10 & $0.61$ & $1418 \pm 27.3 $ & $97.3 \pm 2.9$ & $2140 \pm 22$ &  11 July 2016 & \cite{Alam:2016hwk} \\
11 & $0.32$ & $980.3 \pm 15.9$ & $78.4 \pm 2.3$ & $1270 \pm 14$ &  11 July 2016 & \cite{Alam:2016hwk} \\
12 & $0.57$ & $1387.9 \pm 22.3$ & $96.6 \pm 2.4$ & $2033 \pm 21$ &  11 July 2016 & \cite{Alam:2016hwk} \\
13& $0.31$ & $931.42 \pm 48$ & $78.3 \pm 4.7$ & $1208.36 \pm 33.81$ &  6 December 2016   & \cite{Zhao:2016das} \\
14 & $0.36$ & $1047.04 \pm 44$ & $77.2 \pm 5.7$ & $1388.36 \pm 55$ &  6 December 2016  & \cite{Zhao:2016das} \\
15 & $0.40$ & $1131.34 \pm 44$ & $79.72 \pm 4.9$ & $1560.06 \pm 40$ &  6 December 2016  & \cite{Zhao:2016das} \\
16 & $0.44$ & $1188.78 \pm 32$ & $80.29 \pm 3.4$ & $1679.88 \pm 35$ &  6 December 2016  & \cite{Zhao:2016das} \\
17 & $0.48$ & $1271.43 \pm 25.8$ & $84.69 \pm 3.4$ & $ 1820.44 \pm 39$ &  6 December 2016  & \cite{Zhao:2016das} \\
18 & $0.52$ & $1336.53 \pm 39$ & $91.97 \pm 7.5$ & $1913.54 \pm 47$ &  6 December 2016  & \cite{Zhao:2016das} \\
19 & $0.56$ & $1385.47 \pm 30.5$ & $97.3 \pm 7.9$ & $ 2001.91 \pm 51$ &  6 December 2016  & \cite{Zhao:2016das} \\
20 & $0.59$ & $1423.43 \pm 44$ & $97.07 \pm 5.8$ & $2100.43 \pm 48$ &  6 December 2016  & \cite{Zhao:2016das} \\
21 & $0.64$ & $1448.81 \pm 69$ & $97.70 \pm 4.8$ & $2207.51 \pm 55$ &  6 December 2016  & \cite{Zhao:2016das} \\
22 & $2.33$ & $1669.7 \pm 96.1$ & $224 \pm 8$ & - &  27 March 2017  & \cite{Bautista:2017zgn} \\
23 & $1.52$ & - & - & $3843 \pm 147$ &  16 October 2017  & \cite{Ata:2017dya} \\
24 & $0.81$ & $1586.7 \pm 63.5$ & - & - &  17 December 2017  & \cite{Abbott:2017wcz} \\
25 & $0.72$ & - & - & $2353 \pm 63$ &  21 December 2017 & \cite{Bautista:2017wwp} \\
26 & $1.52$ & $1850 \pm 110$ & $162 \pm 12$ & $3985.2 \pm 162.4$ &  8 January 2018 & \cite{Gil-Marin:2018cgo} \\
27 & $0.978$ & $1586.18 \pm 284.93$ & $113.72 \pm 14.63$ & $2933.59 \pm 327.71$ &  16 January 2018 & \cite{Zhao:2018jxv} \\
28 & $1.230$ & $1769.08 \pm 159.67$ & $131.44 \pm 12.42$ & $3522.04 \pm 192.74$ &  16 January 2018 & \cite{Zhao:2018jxv} \\
29 & $1.526$ & $1768.77 \pm 96.59$ & $148.11 \pm 12.75$ & $3954.31 \pm 141.71$ &  16 January 2018 & \cite{Zhao:2018jxv} \\
30 & $1.944$ & $1807.98 \pm 146.46$ & $172.63 \pm 14.79$ & $4575.17 \pm 241.61$ &  16 January 2018 & \cite{Zhao:2018jxv} \\
\hline
\end{longtable}

\begin{longtable}{ | c | c | c | c| c | }
\caption{The $H(z)$ data compilation presented in Ref. \cite{Jesus:2017zej} and used in the present analysis.} 
\label{tab:data-hz}\\
\hline
   Index & $z$ & $H(z)$ (km/sec Mpc) & $\sigma_{H}$  & Reference\\
\hline   
1 & $0.070$ & $69$    &	$19.6$ & \cite{Zhang:2012mp} \\
2 & $0.090$ & $69$    &	$12$     &	\cite{Simon:2004tf} \\
3 & $0.120$ &	$68.6$ &	$26.2$  &	\cite{Zhang:2012mp} \\
4 & $0.170$&	$83$     &	$8$ &	\cite{Simon:2004tf} \\
5 & $0.179$ &	$75$ &	$4$&	\cite{Moresco:2012jh} \\
6 & $0.199$&	$75$&	$5$&	\cite{Moresco:2012jh} \\
7 & 0.200&	72.9&	29.6&	\cite{Zhang:2012mp} \\
8 & 0.240&	79.69&	6.65&	\cite{Gaztanaga:2008xz} \\
9 & 0.270&	77&	14&	\cite{Simon:2004tf} \\
10 & 0.280&	88.8&	36.6&	\cite{Zhang:2012mp} \\
11 & 0.300&	81.7&	6.22&	\cite{Oka:2013cba} \\
12 & 0.350&	82.7&	8.4&	\cite{Chuang:2012qt} \\
13 & 0.352&	83&	14&	\cite{Moresco:2012jh} \\
14 & 0.3802&	83&	13.5&	\cite{Moresco:2016mzx} \\
15 & 0.400&	95&	17&	\cite{Simon:2004tf} \\
16 & 0.4004&	77&	10.02&\cite{Moresco:2016mzx} \\
17 & 0.4247&	87.1&	11.2&	\cite{Moresco:2016mzx} \\
18 & 0.430&	86.45&	3.68&	\cite{Gaztanaga:2008xz} \\
19 & 0.440&	82.6&	7.8&	\cite{Blake:2012pj} \\
20 & 0.4497&	92.8&	12.9&	\cite{Moresco:2016mzx} \\
21 &0.4783&	80.9&	9&	\cite{Moresco:2016mzx} \\
22 & 0.480&	97&	62&	\cite{Stern:2009ep} \\
23 & 0.570&	92.900&	7.855&\cite{Anderson:2013oza} \\
24 & 0.593&	104&	13&	\cite{Moresco:2012jh} \\
25 & 0.6&		87.9&	6.1&	\cite{Blake:2012pj} \\
26 & 0.68&		92&	8&	\cite{Moresco:2012jh} \\
27 & 0.73&		97.3&	7.0&	\cite{Blake:2012pj} \\
28 & 0.781&	105&	12&	\cite{Moresco:2012jh} \\
29 & 0.875&	125&	17&	\cite{Moresco:2012jh} \\
30 & 0.88&		90&	40&	\cite{Stern:2009ep} \\
31 & 0.9&		117&	23&	\cite{Simon:2004tf} \\
32 & 1.037&	154&	20&	\cite{Moresco:2012jh} \\
33 & 1.300&	168&	17&	\cite{Simon:2004tf} \\
34 & 1.363&	160&	22.6&	\cite{Moresco:2015cya}  \\
35 & 1.43&		177&	18&	\cite{Simon:2004tf} \\
36 & 1.53&		140&	14&	\cite{Simon:2004tf} \\
37 & 1.75&		202&	40&	\cite{Simon:2004tf} \\
38 & 1.965&	186.5&	50.4&	\cite{Moresco:2015cya} \\
39 & 2.300&	224&	8&	\cite{Busca:2012bu} \\
40 & 2.34&	222&	7&	\cite{Delubac:2014aqe} \\
41 & 2.36&	226&	8&	\cite{Font-Ribera:2013wce} \\
\hline
\end{longtable}

\begin{longtable}{ | c | c | c | c| }
\caption{The JLA binned data used in the analysis from Ref. \cite{Betoule:2014frx}}
\label{tab:data-jla}\\
\hline
   Index & $z$ & $\mu$ & $\sigma_{\mu}$ \\
\hline   
1 & $0.01$    &  $32.9539$  & $0.145886$ \\
2 & $0.012$ &  $33.879$  & $0.167796$ \\
3 & $0.014$ &  $33.8421$  & $0.0784989$ \\
4 & $0.016$ &  $34.1186$  & $0.0723539$ \\
5 & $0.019$ &  $34.5934$  & $0.0854606$ \\
6 & $0.023$ &  $34.939$  & $0.0561251$ \\
7 & $0.026$ &  $35.2521$  & $0.0610683$ \\
8 & $0.031$ &  $35.7485$  & $0.0567639$ \\
9 & $0.037$ &  $36.0698$  & $0.0567956$ \\
10 & $0.043$ &  $36.4346$  & $0.0751431$ \\
11 & $0.051$ &  $36.6511$  & $0.0929013$ \\
12 & $0.06$ &  $37.158$  & $0.0620892$ \\
13 & $0.07$ &  $37.4302$  & $0.0658793$ \\
14 & $0.082$ &  $37.9566$  & $0.0546505$ \\
15 & $0.097$ &  $38.2533$  & $0.0599337$ \\
16 & $0.114$ &  $38.6129$  & $0.0374341$ \\
17 & $0.134$ &  $39.0679$  & $0.0386141$ \\
18 & $0.158$ &  $39.3414$  & $0.0346886$ \\
19 & $0.186$ &  $39.7921$  & $0.0321403$ \\
20 & $0.218$ &  $40.1565$  & $0.0329616$ \\
21 & $0.257$ &  $40.565$  & $0.0317198$ \\
22 & $0.302$ &  $40.9053$  & $0.0392622$ \\
23 & $0.355$ &  $41.4214$  & $0.0335758$ \\
24 & $0.418$ &  $41.7909$  & $0.0415207$ \\
25 & $0.491$ &  $42.2315$  & $0.0393713$ \\
26 & $0.578$ &  $42.617$  & $0.0359453$ \\
27 & $0.679$ &  $43.0527$  & $0.0627778$ \\
28 & $0.799$ &  $43.5042$  & $0.0545914$ \\
29 & $0.94$ &  $43.9726$  & $0.0668276$ \\
30 & $1.105$ &  $44.5141$  & $0.154604$ \\
31 & $1.3$ &  $44.8219$  & $0.138452$ \\
\hline
\end{longtable} 

\end{widetext}

\raggedleft
\bibliography{Bibliography}

\end{document}